\documentclass[11pt]{article}
\usepackage{caption}
\usepackage{scalerel}
\usepackage{amsthm,latexsym,amsxtra,graphicx,appendix,epstopdf,feynmf,hyperref,setspace,fix-cm}
\usepackage[normalem]{ulem}
\usepackage{makeidx}
\usepackage{fullpage}
\usepackage{amsmath}
\usepackage{amssymb}
\usepackage{setspace}
\usepackage{bbm}
\usepackage{dsfont}
\usepackage{graphics}
\usepackage{epsfig}
\usepackage[font=footnotesize,labelfont=bf,justification=centerlast,width=.94\textwidth]{caption}
\usepackage{cite}
 \usepackage{multirow}
\usepackage{array,booktabs}
\usepackage{subfigure}
\usepackage{color}
\usepackage[makeroom]{cancel}
\usepackage{tikz}
\usepackage{pgfplots}
\usetikzlibrary{intersections, pgfplots.fillbetween}
\usepackage{tensor}
\usepackage{simplewick}
\usepackage{frcursive}
\usepackage{caption}
\usepackage{hyperref}

\hypersetup{
    bookmarks=true,%
    colorlinks,%
    citecolor=blue,%
    filecolor=blue,%
    linkcolor=blue,%
    urlcolor=blue
}
\usepackage{float}
\usepackage{slashed}
\usepackage{tikz}
\usetikzlibrary{decorations.pathmorphing}

\newcommand{\bi}{\begin{itemize}}
\newcommand{\ei}{\end{itemize}}
\newcommand{\bea}{\begin{eqnarray}}
\newcommand{\eea}{\end{eqnarray}}
\newcommand{\be}{\begin{equation}}
\newcommand{\ee}{\end{equation}}
\newcommand{\dd}{\text{d}}

\newcommand{\tpsi}{\tilde{\psi}}

\newcommand{\tvarphi}{\widetilde{\varphi}}
\newcommand{\snabla}{\slashed{\nabla}}

\newcommand{\btpsi}{\overline{\widetilde{\psi}}}

\newcommand{\la}{\ell_{\scaleto{\text{AdS}}{4.5pt}}}

\def\XXint#1#2#3{{\setbox0=\hbox{$#1{#2#3}{\int}$}
     \vcenter{\hbox{$#2#3$}}\kern-.5\wd0}}

\numberwithin{equation}{section}
\allowdisplaybreaks  

\begin{document}

\vspace*{2.5cm}
\begin{center}
{ \Large \textsc{Finite Features of Quantum De Sitter Space}} \\ \vspace*{1.3cm}
\end{center}

\begin{center}
Dionysios Anninos,$^{1}$ Dami\'an A. Galante,$^{1}$ and Beatrix M\"uhlmann$^{2}$  \\ 

\end{center}
\vskip4mm
\begin{center}
{
\footnotesize
\hspace{1cm}{$^{1}$Department of Mathematics, King's College London, Strand, London WC2R 2LS, UK \newline\newline
$^{2}$ Department of Physics, McGill University, Montreal, QC H3A 2T8, Canada
\\
}}
\end{center}
\begin{center}
{\textsf{\footnotesize{
dionysios.anninos@kcl.ac.uk, damian.galante@kcl.ac.uk, beatrix.muehlmann@mcgill.ca}} } 
\end{center}

\vspace*{0.5cm}

\vspace*{1.5cm}
\begin{abstract}
\noindent
\\ \\ 

We  consider degrees of freedom for a quantum de Sitter spacetime.  The problem is studied from both a Lorentzian and a Euclidean perspective. From a Lorentzian perspective, we compute dynamical properties of the static patch de Sitter horizon. 
These are compared to dynamical features of a black hole. We point out differences suggestive of non-standard thermal behaviour for the de Sitter horizon. We establish that geometries interpolating between an asymptotically AdS$_2 \times S^2$ space  and a dS$_4$ interior are compatible with the null energy condition, albeit with a non-standard decreasing radial size of  $S^2$.  The putative holographic dual of an asymptotic AdS$_2$ spacetime is comprised of a finite number of underlying degrees of freedom.  From a  Euclidean perspective we consider the gravitational path integral for fields over compact manifolds. In two-dimensions, we review Polchinski's BRST localisation of Liouville theory and propose a supersymmetric extension of timelike Liouville theory which exhibits supersymmetric localisation. We speculate that localisation of the Euclidean gravitational path integral is a reflection of a finite number of degrees of freedom in a quantum de Sitter universe.



\end{abstract}




\newpage

\tableofcontents

 \section{Introduction}
 A positive cosmological constant $\Lambda$ such as the one dominating the current epoch of our universe, causes spacetime to expand at an accelerated pace. At sufficiently late times local observers are surrounded by a de Sitter event horizon. The microphysical properties of this horizon and its associated degrees of freedom remain open questions. Of particular interest is the hypothesis of Gibbons and Hawking \cite{GH_1}, based on considerations of black hole thermodynamics, that the cosmological horizon carries an entropy $\mathcal{S}$ whose value is given by the area of the horizon divided by $4G_{\text{Newton}}$ to leading order in the semiclassical approximation. For four-dimensional general relativity endowed with a positive $\Lambda$ one finds
\begin{equation}\label{SdS}
\mathcal{S} \approx  \frac{3\pi c^3}{\Lambda \hbar G_{\text{Newton}}}~.
\end{equation}
The above expression links the putative information content of a de Sitter spacetime to the fundamental constants of cosmology, quantum mechanics, relativity, and gravity. Progress toward a microphysical understanding of $\mathcal{S}$ will provide a novel window toward understanding quantum properties of cosmology. Motivated by (\ref{SdS}), the goal of this note is  to examine the degrees of freedom in a quantum de Sitter space \cite{Banks:2003cg,Banks_dS,Erik_dS,Witten:2001kn, Bousso:2000nf, Fischler_talk,Banks:2001yp,Banks:2018ypk,Banks:2020zcr,Coleman:2021nor,Dio_Frederik_HS} from a variety of complementary perspectives.
Our discussion is partitioned into three main parts. The first of these, section \ref{sec1},  focuses on Euclidean methods, while the second part, section \ref{sec2}, focuses on Lorentzian methods. The final part, section  \ref{sec3}, is more speculative in nature, and aims to tie the considerations of the previous two sections. We now delineate the content of each section in a little more detail.
\newline\newline
In section \ref{sec1}, we consider the properties of $\mathcal{S}$ from the perspective of Euclidean quantum gravity, taking cue from Gibbons and Hawking's proposal \cite{GH_2} that $\mathcal{S}$ is computed, at least semiclassically, by a Euclidean path integral over the sphere. We consider the Euclidean path integral $\mathcal{Z}$ for gravitational theories with positive cosmological constant $\Lambda$, further coupled to matter fields, placed on a compact manifold. Despite its virtues, computing $\mathcal{Z}$ bears several subtleties. For instance, given that the Euclidean gravitational action is unbounded from below one must somehow complexify the path integration contour for the Weyl mode \cite{GHP}. We point out that $\mathcal{Z}$ has exhibits divergences when the theory is placed on compact manifolds whose topology has non-contractible cycles. We examine the properties of such divergences through detailed results in two-spacetime dimensions (\ref{ZCFT_unitary}), as well as  a minisuperspace treatment in three-spacetime dimensions (\ref{msZ}). We observe that these divergences are associated to the infinite dimensionality of the quantum field theoretic Hilbert space, and need to be resolved to fully make sense of the gravitational path integral.
\newline\newline
In section \ref{sec2} we consider properties of the de Sitter horizon from a Lorentzian perspective. We compare and contrast dynamical probes of the de Sitter horizon to those of black hole horizons by studying the scalar field two-point function. We observe that at large timelike separations the two-point function can exhibit oscillatory (\ref{2pti}) or even  non-decaying (\ref{timelike2pt}) behaviour, which contrasts with the exponentially decaying behaviour characteristic of two-point functions in a black hole background. Analytic expressions in the large mass limit are provided, and the interacting case is commented on. Further to this, we consider embedding a portion of the static patch of dS$_4$ into an asymptotically AdS$_2\times S^2$ geometry, and show that these interpolating spacetimes, such as (\ref{flow_geometry}) with (\ref{egflow}), can be supported by null energy condition obeying matter. A timelike AdS$_2$ boundary provides a potential set-up to study the microphysical structure of the de Sitter horizon from the perspective of AdS$_2$ holography. A quantum mechanical, rather than quantum field theoretic,  dual theory comprised of a finite number of microphysical degrees of freedom is envisioned.
\newline\newline
The considerations of sections \ref{sec1} and \ref{sec2} invite us to consider mechanisms that reduce the effective number of degrees of freedom describing a de Sitter spacetime. In section \ref{sec3}, which of a more speculative spirit, we consider such a mechanism at the level of the gravitational path integral $\mathcal{Z}$. Concretely, we introduce a novel two-dimensional theory of quantum gravity described in gauge-fixed form by the action (\ref{SN2sl}). The theory is supersymmetric with $\Lambda>0$, and admits a Euclidean de Sitter saddle. We infer that the theory's gravitational path integral exhibits a form of localisation. We interpret this as a reflection, at the level of the Euclidean path integral, of an underlying finite theory.

\section{Gravitational path integrals}\label{sec1}

%
%

In this section, we study and elaborate on divergences of the Euclidean gravitational path integral for theories placed on  non-trivial topology, and in particular topologies with non-contractible cycles. One hint for the inclusion of non-spherical topologies \cite{Susskind:2021dfc,higher_genus} stems from the relation of the path integral to random matrix theory in two-dimensional quantum gravity \cite{QG_MI1, QG_MI2,QG_MI3}. The path integral on non-spherical topologies relates divergent features of the Gibbons-Hawking path integral to the Hilbert space dimensionality of the quantum matter theory coupled to gravity. 
\newline\newline
\textbf{Dimension one.} The simplest indication that any naive attempt to integrate over Euclidean metrics on a given topology will lead to trouble is the case of a single dimension. Here, the object of interest is a path integral over an einbein $e(\tau)$ with $\tau \sim \tau+1$: 
\begin{equation}\label{beta}
\mathcal{Z}_{S^1} =\int  \frac{\mathcal{D} e(\tau)}{\text{vol} \, \mathcal{G}_1}  e^{-\Lambda \int_{{S^1}} e} =\int_0^\infty  \frac{\dd\beta}{2\beta} e^{-\Lambda \beta} = \frac{1}{2} \int_{\mathbb{R}} \dd\varphi e^{-\Lambda  \,e^{\varphi}}~,
\end{equation}
where we must divide by the volume of the group of one-dimensional diffeomorphisms $\mathcal{G}_1$, and we take the cosmological constant $\Lambda$ to be positive throughout our discussion. In the second equality, we have gauge fixed $e(\tau) = \beta$ and integrated over all $\beta \ge 0$. The variable $\varphi = \log \beta$ is the Weyl factor of the metric in the gauge where the metric is $\tau$ independent. The answer is infinite and a regularisation procedure is required to avoid an infinity from the configuration space of circles of parametrically small proper size. The divergence is logarithmic when expressed in terms of a small $\beta$ cutoff, and stems from the regime $\varphi \lesssim -\log \Lambda/\Lambda_{\text{u.v.}}$, with $\Lambda_{\text{u.v.}}$ some reference cutoff scale. Nonetheless, regardless of the regularisation procedure at small $\beta$, the gauge symmetry of the original problem {is} sufficient to reduce the original expression to an ordinary integral (\ref{beta}). The reduction from the space of paths to the set of non-negative numbers $\beta$ is a simple example of what one can broadly refer to as {localisation}. 

Adding quantum mechanical matter to the problem leads to the more general expression
\begin{equation}
\mathcal{Z}_{S^1} = \int_0^\infty  \frac{\dd\beta}{2\beta} e^{-\Lambda \beta} Z_{\text{QM}}[\beta]~.
\end{equation}
The presence of the thermal partition function $Z_{\text{QM}}[\beta]$ will not cure the small $\beta$ divergence, unless we permit unusual physical conditions such as states of negative norm \cite{Diego_Dio_Stathis} or deform the contour of integration for $\beta$. In fact, even if the dimension of the quantum mechanical Hilbert space $\dim \mathcal{H}_{\text{QM}}$ is finite such that $\lim_{\beta \to 0^+} Z_{\text{QM}}[\beta] =  \dim \mathcal{H}_{\text{QM}}$, the logarithmic divergence in (\ref{beta}) would persist
\begin{equation}\label{logQM}
\mathcal{Z}_{S^1} \approx \dim \mathcal{H}_{\text{QM}} \log \frac{\Lambda_{\text{u.v.}}}{\Lambda}~,
\end{equation}
but now with a modified coefficient related to $\dim \mathcal{H}_{\text{QM}}$. For quantum mechanical theories with $\text{dim}  \mathcal{H}_{\text{QM}} = \infty$, such as the quantum harmonic oscillator, we have a power law divergence instead.
\newline\newline
\textbf{Dimension two.} One might wonder if such phenomena occur in higher dimensions or whether somehow they are a feature of the one-dimensional world. The answer is generally unknown, but there are indications that in two-dimensions a similar phenomenon occurs. On a compact surface $\Sigma_h$ of given genus $h$, the Euclidean gravitational path integral is given by
\begin{equation}\label{2dgrav}
\mathcal{Z}_{\Sigma_h} = e^{\vartheta \chi_h} \int  \frac{\mathcal{D} g_{ij}}{\text{vol} \, \mathcal{G}_2}  e^{-\Lambda \int_{\Sigma_h} \sqrt{g}}~,
\end{equation}
where $\chi_h\equiv 2-2h$ is the Euler characteristic and $\vartheta$ is a dimensionless coupling associated to genus that plays the role of the inverse Newton constant in two-dimensions. 

The path-integral $\mathcal{Z}_{\Sigma_h}$ exhibits divergences for surfaces of small proper area for $h=0,1$ \cite{David,Distler_Kawai}. As we shall review below, adding matter fields to the theory may ease some of the divergences for $h=0$. In addition, unlike the case of one-dimension, the diffeomorphism group is in general not sufficiently large to localise the path integral (\ref{2dgrav}) onto an ordinary integral. 
Nevertheless, for specific choices of matter fields one can propose a definition for $\mathcal{Z}_{\Sigma_h}$.  Adding quantum field theoretic matter generalises (\ref{2dgrav}) to
\begin{equation} \label{Zh}
\mathcal{Z}_{\Sigma_h} =  e^{\vartheta \chi_h} \int  \frac{\mathcal{D} g_{ij}}{\text{vol} \, \mathcal{G}_2}  e^{-\Lambda \int_{\Sigma_h} \sqrt{g}} \, Z_{\text{matter}}[g_{ij}]~.
\end{equation}
In Weyl gauge the path integral (\ref{2dgrav}) becomes a path integral over the Liouville action (see appendix \ref{Liouville_primer} for details).
There exists remarkable evidence that if the matter theory is a minimal model of central charge $c_{\text{m}} < 1$, (\ref{Zh}) indeed localises onto an integral over Hermitian $N\times N$ matrices of large size weighted by a finely tuned matrix potential \cite{QG_MI1, QG_MI2,QG_MI3}. Further to this, quantities computed by the path integral (\ref{Zh}) and expectation values thereof are associated to topological invariants of two-dimensional compact surfaces -- the intersection numbers on certain compactifications of moduli space \cite{Kontsevich,Witten_IT}. Here, the theory is effectively localised onto a discrete set of numbers which are moreover topological invariants. The topological nature of the result might have been anticipated, we are after all trying to integrate  out the Riemannian structure. 

For a matter conformal field theory of central charge $c_{\text{m}}< 1$, it is argued in \cite{kpz,David,Distler_Kawai} that the non-analytic dependence of $\mathcal{Z}_{\Sigma_h}$ on $\Lambda$ is 
\begin{equation}\label{Zh_2d}
\mathcal{Z}_{\Sigma_h} = e^{\vartheta \chi_h} \left(\frac{\Lambda}{\Lambda_{\text{u.v.}}} \right)^{\frac{1}{24}\left(-c_{\text{m}}+\sqrt{(1-c_{\text{m}})(25-c_{\text{m}})}+25\right)\chi_h} f_h(c_{\text{m}})~, \quad\quad h \neq 1~,
\end{equation}
where $f_h(c_\text{m})$ is some real-valued function of $c_{\text{m}}$ for each $h$. We note that the absence of a dimensionful Planck scale in two-dimensions requires the inclusion of a new ultraviolet scale $\Lambda_{\text{u.v.}}$ to make sense of the problem. For $c_{\text{m}}\rightarrow -\infty$ the expression (\ref{Zh_2d}) simplifies to
\begin{equation}\label{8.21}
\mathcal{Z}_{\Sigma_h} \approx e^{\vartheta \chi_h}\left({\frac{\Lambda}{\Lambda_{\text{u.v.}}}}\right)^{-\frac{c_{\text{m}}\chi_h}{12}}~, \quad\quad h \neq 1~.
\end{equation}
On the torus, corresponding to $h=1$, the dependence on $\Lambda$ becomes logarithmic \cite{BK1}. It is tempting to view the coefficient of the logarithm, much like in the one-dimensional case, as a count of the number of states in the theory \cite{Kutasov_Seiberg}. For the series $(m,m+1)$ of unitary minimal models with central charge $c_{\text{m}}= 1-6/(m(m+1))$, (\ref{Zh}) takes the form
\begin{equation} \label{logL}
\mathcal{Z}_{\Sigma_1} \approx \frac{(m-1)}{48} \log \frac{\Lambda_{\text{u.v.}}}{\Lambda}~,
\end{equation}
where, much like for (\ref{logQM}), the logarithmic term stems from the region where the Weyl factor $\varphi$ of the physical metric $g_{ij}$ is $\varphi \lesssim -\log \Lambda/\Lambda_{\text{u.v.}}$. Thus, when the matter theory is a minimal model, the coefficient of the logarithm in (\ref{logL}) can be viewed as a measure of the number of states in the theory coupled to gravity whereby the Virasoro descendants are gauged.\footnote{This is not entirely precise due to an infinite dimensional BRST cohomology \cite{Lian:1991gk}, but it is unclear whether the states in this cohomology which persist even for $c_{\text{m}} = 0$ correspond to physical states.} In one way or another, it seems that conformal field theories with a finite number of Virasoro primaries, such as the minimal models, that are coupled to two-dimensional gravity are the two-dimensional counterpart of a quantum mechanical theory with a finite Hilbert space coupled to one-dimensional gravity. 

When the matter theory takes the form of a two-dimensional conformal field theory of central charge $c_{\text{m}}>1$, the situation is considerably less clear. In particular, in this regime no microscopic completion is known. At genus zero, the path integral (\ref{Zh}) for $c_{\text{m}} \gg 1$ can be computed in a sensible way using a semiclassical path integral representation of timelike Liouville theory \cite{timelike, Muhlmann:2022duj,Giribet:2022cvw}. In contrast to $c_{\text{m}}<1$, the gravitational path integral of timelike Liouville theory on the two-sphere is finite for small physical area upon dividing by a residual volume of $PSL(2,\mathbb{C})$. At genus one, we can consider the general form of the torus path integral of a two-dimensional conformal field theory. The asymptotic growth of states is captured by the high temperature behaviour, which corresponds to a torus of complex structure $\tau = \tau_1+ i \tau_2$ with $\tau$ approaching the origin of the complex $\tau$-plane. In this limit, the partition function of a unitary conformal field theory of central charge $c_{\text{m}}$ scales according to the Cardy behaviour \cite{Cardy}
\begin{equation}\label{cardy}
\lim_{\tau_2 \to 0^+} Z_{\text{CFT}}[  i \tau_2] \approx   e^{\frac{\pi c_{\text{m}}}{6\tau_2}}~. 
\end{equation}
If we select the fundamental domain to contain the origin, the essential singularity at $\tau_2$ will dominate the integration over moduli space. Incorporating the effect of $\mathfrak{b}\mathfrak{c}$-ghosts and the Liouville field leads to the more exact expression
\begin{equation}\label{T2div}
 \mathcal{Z}_{\Sigma_1} = {\color{black}  \pm \frac{i}{2{\color{black}\sqrt{2}\pi}\beta}} \log \frac{\Lambda}{\Lambda_{\text{u.v.}}} \times \int_{\mathcal{F}} \frac{\dd^2 \tau}{\tau_2^{3/2}}  \sum_{\Delta,\bar{\Delta}} q^{\Delta-\frac{c_{\text{m}}}{24}} \bar{q}^{\bar{\Delta}-\frac{{c}_{\text{m}}}{24}}~,
\end{equation}
where $q \equiv e^{2\pi i \tau}$. The summation is over the Virasoro primaries of the $c_{\text{m}}>1$ matter CFT for which we have assumed there are no null states. The derivation of (\ref{T2div}) follows  \cite{BK1}, and the overall phase in (\ref{T2div}) comes from the wrong sign kinetic term in timelike Liouville theory. Due to the Cardy behaviour (\ref{cardy}), expression (\ref{T2div}) diverges exponentially in $1/\tau_2$ near the origin of the complex $\tau$-plane which we have included in $\mathcal{F}$. This divergence is the two-dimensional counterpart of a quantum mechanical theory of infinite Hilbert space coupled to one-dimensional gravity.  

At higher genus, and using the approach of the sewing formula  \cite{Sen:1990bt}, an analogous divergence will manifest itself in those regions of moduli space where we have tubes of length $\sim 1/\tau_2$ connecting the multiple three-punctured two-spheres. For instance, at genus two the CFT partition function reads
\begin{equation}\label{ZCFT_unitary}
Z_{\text{CFT}}[\upsilon;\tau] = \left(\frac{\upsilon}{\upsilon_0}\right)^{-\frac{c_{\text{m}}}{6}} \sum_{i,j,k} |C_{ijk}|^2 q^{\left( \Delta_i+\Delta_j+\Delta_k - \frac{c_{\text{m}}}{8} \right)} \bar{q}^{\left( \bar{\Delta}_i+ \bar{\Delta}_j+\bar{\Delta}_k - \frac{{c}_{\text{m}}}{8}\right)}~,
\end{equation}
where $\upsilon$ is the physical area of $\Sigma_h$, and $\upsilon_0$ is a reference area. The $C_{ijk}$ are OPE coefficients of three states on the punctured two-sphere and the sum runs over the entire Hilbert space.  Provided the lowest weight operator is the identity, one can use modular properties to show that the above expression diverges as $e^{{\pi c_{\text{m}}}/{(2\tau_2)}}$. For genus $h$, similar reasoning leads us to postulate that the divergence will grow as
\begin{equation}\label{genush}
\lim_{\tau_2 \to 0^+} Z_{\text{CFT}}[\upsilon;i\tau_2] \approx  \left(\frac{\upsilon}{\upsilon_0}\right)^{\frac{ c_{\text{m}}}{12}\chi_h} e^{-\frac{ \pi c_{\text{m}}}{4\tau_2}\chi_h}~, \quad\quad h \ge 2~.
\end{equation} 
The divergence stems from the region in moduli space where the $2h-2$ punctured two-spheres are connected by $3h-3$ tubes of length $\sim 1/\tau_2$. In string theory, such a divergence is associated with tachyonic states in the target space. Here, we interpret it as a divergence associated to the infinite-dimensional quantum field theoretic Hilbert space. 

For  $c_{\text{m}} \ge 25$, the small $\tau_2$ divergence  appearing at genus $h \ge 1$ cannot be cured by coupling the theory to two-dimensional quantum gravity \cite{Kutasov_Seiberg,Bautista:2019jau}.
 To make sense of the problem, it would seem we must invoke a mechanism that vastly reduces the contributions from the infinite dimensional field theoretic state space. One possibility is to permit unusual physical conditions such as states of negative norm \cite{Diego_Dio_Stathis}, or more generally some grading of states in the trace that permits sign changes as happens for the superstring torus partition function.\footnote{One might imagine trying to first sum over all the genera and then performing the integrals over moduli space. Such a possibility, though perhaps of some interest, would require additional principles to be justified. 
}  Alternatively, we could also consider deforming the contour of integration for $\tau_2$ into the complexified $\tau_2$ plane -- this approach will invariably present further ambiguities. Given the divergences at higher genera for $c_{\text{m}}\ge 25$, it is interesting to note that a potentially ultraviolet finite quantity independent of the normalisation of the path-integral is given by the ratio of the sphere path integral to the square of the disk partition function (this quantity was investigated in \cite{Mahajan:2021nsd} for $c_{\text{m}} \ll 1$ and the disk partition function of timelike Liouville theory was studied in \cite{Bautista:2021ogd}).
\newline\newline
The general lesson is that much like for one-dimensional quantum gravity, the two-dimensional Euclidean path integral over compact manifolds needs to be regularised in a non-standard way. In particular, the divergences that appear are not related to local ultraviolet divergences of quantum field theory and cannot be absorbed in local counterterms. Rather, they stem from integrating over all geometries resulting in a  sensitivity of the path integral to the infinite dimensional Hilbert space of the matter quantum theory.
\newline\newline
\textbf{Dimension three.} In three-dimensions significantly less is known. Restricting, for simplicity, to an $S^3$ topology we can study the path integral  
\begin{equation}\label{3dgrav}
\mathcal{Z}_{S^3} = \int  \frac{\mathcal{D} g_{ij} }{\text{vol} \, \mathcal{G}_3}  e^{- \frac{1}{16\pi G_3}  \int_{S^3} \sqrt{g} \, \left( R - 2 \Lambda  \right)}~,
\end{equation}
where $\Lambda>0$ is the cosmological constant and $G_3$ is the three-dimensional Newton constant. Taken at face value, the above path integral is infinite. The infinity in this case stems from configurations with rapid spatial variations of the conformal factor. To define the above path integral, an alternative contour over the space of metric configurations has been proposed by Gibbons, Hawking, and Perry \cite{GHP}. The procedure is guided by the presence of a semiclassical saddle at small values of the dimensionless combination $\Lambda^{1/2} G_3$. The round metric on $S^3$ of size $1/\sqrt{\Lambda}$ reveals itself as a real valued saddle. One then complexifies the contour of the unbounded conformal mode fluctuations to a bounded one. 

To compute (\ref{3dgrav}) about the $S^3$ saddle, one can employ the relation of three-dimensional gravity to Chern-Simons theory and argue that loop corrections are captured by an analytic continuation of known results in Chern-Simons theory. For example, up to the one-loop correction we have \cite{Witten_JP, Adams:1996qe}
\begin{equation}\label{3doneloop}
\mathcal{Z}_{S^3} \approx e^{-\frac{\pi i}{2}} e^{\frac{\pi}{2\Lambda^{1/2} G_3}} \, \frac{\left( {8 \Lambda^{1/2} G_3} \right)^{3}}{\text{vol} \, SO(4)}  \left| \frac{\text{det}' \, \Delta }{\sqrt{\text{det} \, \text{d}^\dag \text{d}} } \right|~,
\end{equation} 
where $\Delta$ is the scalar Laplacian on the round $S^3$, $\text{d}^\dag \text{d}$ is the Laplacian acting on transverse one-forms on $S^3$, and the prime indicates we are dropping zero modes in the functional determinants
The absolute value of the ratio of determinants found in (\ref{3doneloop}) is a well known expression for the analytic torsion of a three-manifold which is a topological invariant. 

Both the relation to Chern-Simons theory, whose path integral is known to localise onto a type of matrix integral \cite{Marino_CS,Tierz_CS, Witten_Beasley},
 as well as the appearance of the analytic torsion suggest that if any sense can be made of the path integral (\ref{3dgrav}) it might itself be computing topological invariants. Indeed, it has been argued that the semiclassical expansion of Chern-Simons theory yields topological invariants at each order of the perturbative expansion \cite{Dimofte:2009yn}. 

In addition to $S^3$, Lens spaces also admit saddle point solutions. The Lens spaces constitute an infinite discretuum of compact three-manifolds which are smooth quotients of $S^3$, and have a finite fundamental group. Their contributions to the gravitational path integral, as explored in \cite{Castro:2011xb}, is exponentially suppressed  in $\Lambda^{1/2} G_3$ as compared to $\mathcal{Z}_{S^3}$. The remaining compact three-manifolds do not admit smooth Einstein metrics. 

Less is known about the three-dimensional gravitational path integral on manifolds with a non-contractible cycle. To obtain some idea, we consider the theory on an $S^1 \times S^2$ manifold coupled to $N_s$ free scalar fields.  We moreover consider the problem in a minisuperspace approximation where the metric on $S^1\times S^2$ is given by
\begin{equation}
ds^2 = \dd\tau^2 + L^2 \dd\Omega^2_2~, \quad \quad \tau \sim \tau + \beta~.
\end{equation}
The Euclidean gravitational action on the above configurations evaluates to
\begin{equation}
- S_E =  \frac{\beta}{2 G_3} \left(  1  -   L^2   \Lambda \right)~.
\end{equation}
We see that the minisuperspace action does not suppress parametrically large temperatures. On the other hand, we have for the partition function of a quantum field theory on $S^1 \times S^2$ in the high temperature limit
\begin{equation}\label{highT}
\lim_{\beta/L \to 0^+} Z_{\text{QFT}}[\beta,L] \approx  e^{C  N_s L^2/\beta^2}~,
\end{equation}
with $C$ an order one constant. 
 The behaviour in (\ref{highT}) takes place once the temperature $1/\beta$ is above any characteristic energy scale $\mu$ in the theory. We further assume this occurs well below the Planck scale, i.e. $G_3 \ll \beta \ll 1/\mu$. Thus, at least within the minisuperspace approximation we have
 \begin{equation}\label{msZ}
 \mathcal{Z}_{S^2\times S^1} \approx \int \dd \mu(L,\beta) \, e^{ \frac{\beta}{2 G_3} \left(  1  -   L^2   \Lambda \right)} Z_{\text{QFT}}[\beta,L]~.
 \end{equation}
The partition function $\mathcal{Z}_{S^2\times S^1}$ is dominated by the growth (\ref{highT}) of the integrand at small $\beta$.\footnote{We further note that for $L^2 \Lambda < 1$, the large $\beta$ part of the integral in (\ref{msZ}) is also potentially divergent. This divergence can be cured by suitably tuning the ground state energy $E_0$.}
 Comparing to (\ref{3doneloop}) we see that for
\begin{equation}\label{betacrit}
\frac{\beta^2}{L^2} \approx  {N_s} {G}_3 \Lambda^{1/2}~,
\end{equation}
the partition functions $\mathcal{Z}_{S^2\times S^1}$ and $\mathcal{Z}_{S^3}$ become comparable in the semiclassical limit. The temperature (\ref{betacrit}) can be parametrically smaller than the Planck scale.
\newline\newline
\textbf{Dimension four.} In four-dimensions, there exist no results for the Euclidean gravitational path integral of Einstein gravity with $\Lambda > 0$ beyond one-loop order. As for three-dimensions, the theory exhibits the round metric on the four-sphere as a saddle\footnote{Interestingly, it is not currently known whether other smooth Einstein metrics exist on an $S^4$ topology. This is in stark contrast to the case of odd dimensional spheres in dimension five and above, which are known to be infinitely many already for $S^5$. Other four-dimensional manifolds admitting smooth Einstein metrics are $\mathbb{C}\mathbb{P}^2$, $S^2 \times S^2$ and $\mathbb{C}\mathbb{P}  \# k \overline{\mathbb{C}\mathbb{P}^2}$ with $1 \le k \le 8$.} and one can argue that there is a sensible semiclassical expansion in $\Lambda G_4$ around that configuration. To one-loop \cite{long_sphere,Law}
\begin{equation}\label{4doneloop}
\mathcal{Z}_{S^4} \approx  e^{-\pi i} \, e^{\frac{3\pi}{\Lambda G_4}} \left( \frac{\Lambda G_4}{{3\pi}} \right)^{5} \sqrt{\frac{\text{det}' \left( \Delta_{(1)} -3 \right)}{\text{det} \left( \Delta_{(2)} +2 \right)}}~,
\end{equation} 
where $\Delta_{(1)}$, and $\Delta_{(2)}$ are the Laplacians on transverse vector fields and transverse traceless tensors respectively. 

At first sight, (\ref{3doneloop}) and (\ref{4doneloop})  seem fairly similar. However, there is an important difference one confronts upon trying to evaluate the functional determinants. In three-dimensions all divergences can be absorbed into local counterterms, and a meaningful constant can be obtained from regularising the functional determinants in  (\ref{3doneloop}). On the other hand, the four-dimensional functional determinants have a logarithmic divergence that is sensitive to some reference scale, such as the ultraviolet cutoff scale, that cannot be removed through the addition of local counterterms. Due to this logarithmic term, a topological interpretation of the one-loop contribution to the four-dimensional path integral of pure Einstein gravity with $\Lambda > 0$ seems unclear. Instead, the absence of a topological interpretation for the  one-loop approximation of the gravitational path integral may serve indication that the theory requires a completion. For instance, one could add further particles to the theory arranged in such a way that the coefficient of the logarithmic divergence vanishes. In fact, using the techniques of \cite{long_sphere}, we can see that adding a free Dirac fermion of mass $m_f^2 =  \Lambda(\sqrt{78}+8)/3$ and two minimally coupled free scalars of mass $m_b^2 = \Lambda (2 \sqrt{78}+11)/3$ will lead to an entirely ultraviolet finite expression at one-loop. More elaborate spectra may lead to more elaborate cancellations at increasingly high order in the loop expansion. As a different but somewhat related example, we note that the twisted $\mathcal{N}=2$ four-dimensional topological field theory \cite{Witten:1988ze}  exhibits no logarithmic divergence.
(In two-dimensions the logarithmic divergence is not  present since the combined central charge from the matter, $\mathfrak{b}\mathfrak{c}$-ghost, and gravitational sectors vanishes.)

A somewhat exotic spectrum, studied in the context of higher-spin theories \cite{Vasiliev}, involves an infinite tower of massless Fronsdal gauge fields in four-dimensions. Although there are classical equations for higher spin theory on $S^4$, the corresponding path integral and its ultraviolet completion remain open problems. There is some circumstantial indication that it will localise onto an ordinary matrix type integral due to the vast higher spin gauge symmetry of the underlying theory. This results in the expectation values of higher spin gauge invariant observables being computed by $N\times N$ matrix integrals with $N \sim 1/(\Lambda G_4)$  \cite{Dio_Frederik_HS}. 
\newline
\begin{center}   {\textit{Brief synthesis}}   \end{center}
Our discussion reiterates that we should treat the gravitational path integral with  caution. From the above considerations it exhibits at least two types of divergences. One of these is a divergence due to the unboundedness of the Euclidean action stemming from the wrong sign kinetic term of the conformal mode. One might hope this divergence can be resolved by a judicious choice of integration contour \cite{GHP}, or some other considerations \cite{marolfi}. The second potential divergence, which we studied explicitly in one and two dimensions, and only schematically for higher dimensions, arises for topologies with a product structure $S^1 \times \mathcal{M}$ and stems from the infinite dimensionality of the quantum field theoretic Hilbert space.
\newline\newline
{\textbf{Euclidean thermodynamics.}} From the perspective of the Euclidean picture of black hole thermodynamics, significant evidence has surmounted that the gravitational path integral encodes a rich amount of information about the semiclassical limit of a quantum theory of gravity (recent developments include \cite{Dong:2018cuv,Almheiri:2019qdq}). The gravitational path integral on compact manifolds makes a particularly striking appearance in the context of the semiclassical quantisation of cosmological spacetimes. The basic postulate due to Gibbons and Hawking \cite{GH_2}, which generalises the maxim of black hole thermodynamics to cosmological spacetimes governed by a positive $\Lambda$ term, states that the entropy $\mathcal{S}$ of the cosmological horizon is given by
\begin{equation}\label{SlogZ}
\mathcal{S} = \log \mathcal{Z}~,
\end{equation}
where $\mathcal{Z}$ is the Euclidean gravitational path integral appropriately summed over all compact manifolds. The absence of an energy term in $\mathcal{Z}$, which is meant to be a thermal partition function, stems from the compactness of the Cauchy slices leaving no room to place the  ADM energy term in the Hamiltonian formalism of general relativity. To leading order in the semiclassical limit, the gravitational path integral is dominated by the Euclidean de Sitter geometry  -- the $d$-dimensional round-sphere -- on an $S^d$ topology.

Both $\mathcal{Z}$ as well as the sum over compact manifolds are notions that become increasingly difficult to define in four spacetime dimensions. However, as discussed, one can compute quantities employing the saddle point approximation.  In four-dimensions \cite{long_sphere}
\begin{equation}\label{ghE}
\mathcal{S} = \frac{3\pi}{\Lambda G_4} + 5 \log \Lambda G_4 + \frac{571}{90} \log \Lambda \ell_{\text{ref}}^2 + \mathcal{O}(1)~.
\end{equation}
As we will discuss next, the leading term  in the above expansion is the area of the de Sitter event horizon divided by $4 G_4$, which is the horizon entropy \cite{GH_1} according to the Bekenstein-Hawking relation. 
Along a similar vein, the two-dimensional partition function (\ref{8.21}) at genus zero resembles the partition function of a CFT of central charge $c_{\text{m}}$ on a round two-sphere with Ricci scalar $R= 2\Lambda$ \cite{Zamo_scaling}. This can be viewed as an entanglement entropy for the density matrix obtained by tracing out the modes behind the two-dimensional de Sitter horizon \cite{Holzhey,CC, Casini}. Such considerations allow one to relate the somewhat esoteric Euclidean picture discussed so far to a physical picture in Lorentzian spacetime. 


\section{Lorentzian signature}\label{sec2}

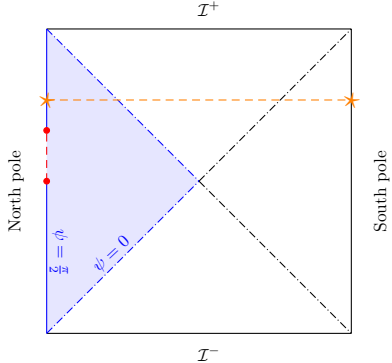
\begin{figure}[H]
\begin{center}
\begin{tikzpicture}[scale=2.7]
   
\node[scale=.6,rotate=90] at (-.15, .7)   {$\text{North pole}$ ~ };
\node[scale=.6,rotate=90] at (1.65, .7)   {$\text{South pole}$ ~ };
\node[scale=.6,rotate=270,blue] at (.06, .4)   {$\psi= \frac{\pi}{2}$ ~ };
  \draw (1.5,0)--(0,0);    
  \draw (1.5,1.5) -- (0,1.5);
    \draw (1.5,0)--(1.5,1.5);    
    \draw[blue] (0,0)--(0,.75);  
    \draw[blue] (0,1)--(0,1.5);   
    \draw[densely dashed,red] (0,.75)--(0,1);        
    \node[scale=.6] at (.8, 1.6)   {$\mathcal{I}^+$ ~ };
  \draw[densely dashdotted] (.75,.75) -- (1.5,1.5);    
 \draw[name path=A, densely dashdotted, blue] (0,0) -- (.75,.75);  
    \draw[name path=B, densely dashdotted,blue] (0,1.5) -- (.75,.75);  
    \tikzfillbetween[of=A and B]{blue, opacity=0.1};
      \draw[densely dashdotted] (.75,.75) -- (1.5,0);   
\node[scale=.6] at (.8, -.1)   {$\mathcal{I}^-$ ~ };
\draw [fill, red] (0, .75) circle [radius=0.014];
\draw [fill, red] (0, 1) circle [radius=0.014];
\node[scale=1.1,rotate=60] at (0, 1.15)   {${\color{orange}\star}$ ~ };
\node[scale=1.1,rotate=60] at (1.5, 1.15)   {${\color{orange}\star}$ ~ };
  \draw[densely dashed,orange] (0, 1.15) -- (1.5, 1.15);    
 \node[scale=.6,rotate=43,blue] at (.32, .39)   {$\psi=0$ ~ };      

\end{tikzpicture}
\end{center}
\caption{\small{Penrose diagram of global (full) and static (shaded) dS. The dots and stars connected by a vertical and horizontal line are indicating examples of timelike and spacelike separated points respectively.}}
\label{fig:penrose}
\end{figure}

In  Lorentzian signature, a classical theory of gravity with positive $\Lambda$ admits a vast configuration space of exponentially expanding solutions \cite{FG,Starobinsky}. According to the cosmic no hair hypothesis (see for example \cite{Wald:1983ky}) within any small neighbourhood the expansion will dilute other matter fields such that the space will become increasingly similar to that of a pure de Sitter universe. Granting the cosmic no hair hypothesis, then, leads to the conclusion that in the vicinity of any local observer the spacetime will resemble the static patch of a de Sitter universe
\begin{equation}\label{sp}
ds^2 = \frac{3}{\Lambda} \left( - \sin^2\psi  \dd t^2+ \dd\psi^2 + \cos^2\psi \, \dd\Omega_2^2 \, \right)~,
\end{equation}
where we will restrict ourselves to four spacetime dimensions for the sake of concreteness (much of the discussion can be generalised). A cosmological event horizon   appears at $\psi  = 0$ of area $A_{\text{hor}} = 12\pi/\Lambda$, while the observer's worldline is at $\psi = \pi/2$, as displayed in figure \ref{fig:penrose}. The metric covering the complete manifold of the four-dimensional de Sitter spacetime is
\begin{equation}\label{global}
ds^2  = \frac{3}{\Lambda}  \left(-\dd T^2 +   \cosh^2 {{T}}  \, \dd\Omega_3^2 \right)~,
\end{equation} 
where we call $T \in \mathbb{R}$ the global time. Along an inertial worldline, this global time coincides with the static-patch time $t$ in (\ref{sp}). 
\noindent
Gibbons and Hawking argue \cite{GH_1} that the cosmological horizon carries an entropy
\begin{equation}\label{gharea}
\mathcal{S}_{\text{hor}} = \frac{A_{\text{hor}}}{4 G_4} = \frac{3\pi}{\Lambda G_4}~,
\end{equation} 
much like its black hole counterpart. To leading order in a small $\Lambda G_4$ expansion, the contribution $\mathcal{S}$ in (\ref{ghE}) to the Euclidean path integral gives $\mathcal{S} = \mathcal{S}_{\text{hor}}$. If indeed the cosmological horizon carries entropy, this agreement suggests that the Euclidean picture, at least semiclassically, can be used as a method to compute the perturbative structure of the horizon entropy. 

From a Lorentzian perspective, computing quantum corrections to the leading expression (\ref{gharea}), which may stem from the entanglement of quantum fields across the de Sitter horizon among other origins, remains elusive. The first logarithmic term in (\ref{ghE}), for instance, has no known Lorentzian counterpart. Nonetheless, according to Gibbons and Hawking it  contributes to the cosmological horizon entropy.\footnote{Similar logarithmic terms have been discussed by Sen \cite{Sen:2012dw} in the context of black holes, and have been schematically associated to the broken translation symmetries due to the location of the black hole. It is unclear whether the logarithmic terms in (\ref{ghE}) are of a similar  origin. Their coefficient stems from the dimensionality of the four-sphere isometry group. In the context of Chern-Simons theory, analogous terms have been argued to contribute to the englement entropy \cite{Kitaev:2005dm,Levin:2006zz}.} 

{\subsection{Degree of freedom counting, naively.} 

We now consider some of the consequences that follow from assuming $\mathcal{S}= \log \mathcal{Z}$ is an entropy. Firstly, since the gravitational path integral over compact manifolds does not fix the size of the thermal circle we should note that the entropy is computed in the microcanonical ensemble. This is consistent with the fact that all configurations have vanishing energy. A horizon whose microstates are all exactly degenerate in energy is unusual, one typically considers contributions from a narrow band of energies. (A notable exception is a BPS black hole whereby supersymmetry enforces that all microstates have degenerate energy.) Moreover, if the spectrum has only vanishing energy configurations and a finite entropy, we are led to the conclusion that the theory has a finite dimensional Hilbert space $\mathcal{H}_{\text{GH}}$ of dimension $\text{dim} \, \mathcal{H}_{\text{GH}} = e^{\mathcal{S}}$ \cite{Banks:2003cg,Banks_dS,Erik_dS,Witten:2001kn, Bousso:2000nf, Fischler_talk,Banks:2001yp,Banks:2018ypk,Banks:2020zcr}, and moreover that the full density matrix is proportional to the identity operator (related discussions can be found in \cite{Banks:2003cg,Dong:2018cuv,Anninos:2021ihe,Chandrasekaran:2022cip}). 

A finite dimensional Hilbert space seems to be in complete contradiction to the semiclassical picture of quantum fields propagating on an approximately rigid spacetime. Here, one finds an infinite dimensional  field theoretic Hilbert space. We can elaborate on this point by imposing an ultraviolet cutoff length $\ell_{\text{u.v.}} \equiv \sqrt{1/\Lambda_{\text{u.v.}}}$. Even if we restrict the volume of a spatial slice to be of the order of the de Sitter length, $\ell \equiv \sqrt{3/\Lambda}$, our underlying theory would have $\mathcal{N}_{\text{u.v.}} \sim  \ell^3/\ell^3_{\text{u.v.}}$ degrees of freedom. If we conservatively endow each of these degrees of freedom two states we end up with a Hilbert space of dimension $\text{dim} \, \mathcal{H}_{\text{u.v.}} \approx 2^{\mathcal{N}_{\text{u.v.}}}$. We observe that
\begin{equation}\label{uvdim}
 \text{dim} \, \mathcal{H}_{\text{u.v.}}  \gg \text{dim} \, \mathcal{H}_{\text{GH}}~. 
\end{equation}
This type of counting problem permits the independent excitation of states, and does not take into account the effects of gravitational backreaction. 

An alternative method for estimating the state space is given by trying to encode all the states in the interior of the cosmological horizon in the form of a black hole. In asymptotically anti-de Sitter space or flat spacetimes, one can tune the size of the black hole to be arbitrarily large from which it follows, assuming the Bekenstein-Hawking relation, that the underlying theory has an infinite dimensional state-space. In de Sitter space there is a bound on the size of a black hole, discovered by Nariai, that forbids an arbitrarily large amount of black hole entropy to be placed within the cosmological horizon. The entropy of a black hole saturating the Nariai bound $\mathcal{S}_{\text{Nariai}} = \mathcal{S}_{\text{hor}}/3$ is less than that of the cosmological horizon of an empty de Sitter universe. From this perspective, the number of states in $\text{dim} \, \mathcal{H}_{\text{GH}}$, though finite, is more than sufficient to support excitations of matter fields within the cosmological horizon. The basic tension in (\ref{uvdim}) is softened by the fact that most states in $\text{dim} \, \mathcal{H}_{\text{u.v.}}$ would lead to significant gravitational backreaction thereby invalidating the naive count. 

The finiteness of the Hilbert space as discussed above from a Lorentzian perspective, is  perhaps also welcome from the Euclidean perspective. For instance, we observed in Euclidean expressions such as (\ref{T2div}) that once additional topologies of the type $S^1 \times \mathcal{M}$ are included there are significant divergences associated to the infinite dimensionality of the field theoretic state-space. To render such divergences (in the Euclidean picture) finite might involve reducing the dimensionality of the field theoretic state-space. 
\newline\newline
The interplay between Euclidean topology, finiteness of the gravitational path integral, and holographic considerations of the Hilbert space dimensionality for theories with $\Lambda > 0$ warrants a deeper understanding. 

\subsection{Dynamical probes on a rigid geometry}

The possibility of a finite dimensional Hilbert space (or more conservatively, a finite number of underlying degrees of freedom) can be further pursued by exploring potential dynamical consequences. To do so, one must  decide what are good probes with which to measure the dynamical effects of the cosmological horizon. This problem is challenging due to the absence of a spatial timelike boundary where the dynamical behaviour of the metric field is under control. 

We first consider the limit of quantum fields on a rigid de Sitter spacetime.
Quantum correlations in the equilibrium state can be obtained by Wick rotating expectation values of local field insertions on the Euclidean sphere. The metric on the sphere can be conveniently realised in two distinct coordinate systems. One is the Euclidean continuation $t \to -i t_E$ of the static patch coordinates (\ref{sp}), yielding
\begin{equation}\label{sphere1}
ds^2 =  \frac{3}{\Lambda}  \left( \sin^2\psi \dd t_E^2 + \dd \psi^2 + \cos^2\psi \dd\Omega_2^2 \right)~, \quad\quad t_E \sim t_E + 2\pi~.
\end{equation}
The other is the standard coordinate system on the round $S^4$, namely
\begin{equation}\label{sphere2}
ds^2 =  \frac{3}{\Lambda}  \left(\dd \theta^2  + \cos^2\theta  \dd\Omega_3^2 \right)~,
\end{equation}
with $\theta \in [-\pi/2,\pi/2]$ obtained as the Euclidean continuation $T\rightarrow -i\theta$ of the global de Sitter patch (\ref{global}). Path integrating fields over the Euclidean sphere with $n$ insertions of some local observable $\mathcal{O}(\mathcal{X})$ on $S^4$ yields Euclidean correlation functions. They become Lorentzian correlations upon analytically continuing $\mathcal{X} \in \Omega_4$, with a specific $i \varepsilon$ prescription, to time ordered points in the Lorentzian de Sitter spacetime. The metrics (\ref{sphere1}) and (\ref{sphere2}) allow us to compute correlations in either the static patch or global de Sitter coordinates. 

As a simple example, we consider the two-point function of a free massive scalar field $\Phi(x)$ with $x\equiv (T,\boldsymbol{\theta})$ and $\boldsymbol{\theta}\in \Omega_3$.  This is known analytically in any number of dimensions, for any mass and any two spacetime points \cite{Spradlin:2001pw,Anninos:2012qw}. In $d=4$, one finds the Wightman two-point function
\begin{equation}\label{2pt}
\langle E | \Phi(x') \Phi(x) | E \rangle = \frac{\Gamma(\Delta) \Gamma(\bar{\Delta})}{16\pi^2} \, _2F_1 \left( \Delta,\bar{\Delta}; 2; \frac{1+P_{x',x}}{2} \right)~,
\end{equation}
where $\bar{\Delta} \equiv 3-\Delta$ and $\Delta\bar{\Delta}$ is the mass squared of $\Phi(x)$ in units of $\ell$. Group theoretically, the quantity $\Delta\bar{\Delta}$ corresponds to the eigenvalue of the quadratic Casimir of the unitary irreducible representation furnished by single particle states built on top of the Euclidean vacuum $|E\rangle$. 
The embedding distance is given by
\begin{equation}\label{embedding_distance}
P_{x',x}\equiv \eta_{IJ} X^{'I} X^J~,\quad \eta_{IJ}= \text{diag}(-1,1,1,1,1)~,
\end{equation}
where $X^0 = \sinh T$ and $X^i= \omega^i\cosh T$; $\omega^i\in \mathbb{R}$ are coordinates on the three-sphere $\omega_i\omega^i=1$, $i= 1,\ldots, 4$.
The two-point function (\ref{2pt}) can also be obtained upon Wick rotating the Euclidean two-point function on $S^4$ to Lorentzian signature.
\newline\newline
{\textbf{Timelike separation.}}
First, we study timelike separated points both sitting at the North pole of $S^3$. This is displayed by the two dots connected by a vertical dashed line in figure \ref{fig:penrose}.  The two-point function behaves as 
\begin{equation}\label{2pti}
\langle E | \Phi(T,\boldsymbol{\theta}_N) \Phi(0,\boldsymbol{\theta}_N) | E \rangle \approx \begin{cases}
 \alpha (\Delta) e^{-\Delta (T+ i \pi)}  + \alpha (\bar{\Delta}) e^{-\bar{\Delta} (T+ i \pi )}  ~, &\text{for $T \gg 1$ and $\Delta\bar{\Delta}>0$~,}\\ \noalign{\vskip4pt}
-\frac{\sqrt{2} (\Delta \bar{\Delta})^{1/4} }{8\pi ^{3/2} \sinh ^{{3}/{2}}T} e^{-i \left( T \sqrt{\Delta \bar{\Delta}} - \frac{\pi}{4}\right)}\,, & \text{for $\Delta \bar{\Delta} \gg 1$ \,, }
\end{cases} 
\end{equation}
where
\begin{equation}\label{alpha}
\alpha(\Delta) = \frac{2^{2 \Delta -3} \sin (\pi  \Delta ) \Gamma (2-2 \Delta ) \Gamma (\Delta )^2}{\pi ^3} \, ,
\end{equation} 
and the subscript ``N'' on the coordinate indicates that the spherical coordinates $\boldsymbol{\theta}$ reside at the North pole of $S^3$.
Note that $\alpha(\Delta)$ has poles at $\Delta = 1/2 + \mathbb{N}$ that are not present in the original two-point function (\ref{2pt}). These are an artefact of the limit we are taking. 
The correlator (\ref{2pti}) is complex and has an exponentially decaying part both in the long time separation and in the large-mass limit. This reflects the dilution of massive matter due to the expansion of space. 

When $\Delta \bar{\Delta} > 9/4$,  corresponding to heavy fields, there is a piece of the two-point function that oscillates in a way that resembles a non-dissipative normal mode behaviour, see figure \ref{fig:2pt_time}. Moreover, for the complementary series values $\Delta \in (0,3)$ permitted by unitarity, one can suppress the exponential decay of the correlator almost entirely.  In fact, for small $\Delta$ and long times the correlator (\ref{2pt}) decreases linearly with time,
\begin{equation}\label{timelike2pt}
\langle E | \Phi(T,\boldsymbol{\theta}_N) \Phi(0,\boldsymbol{\theta}_N) | E \rangle \approx \frac{3}{8 \pi ^2 \Delta \bar{\Delta}}-\frac{T}{8 \pi ^2} \,, \qquad  \text{for $0<\Delta \bar{\Delta} \ll 1$ and $T \gg 1$} \,. 
\end{equation}
This formula is valid up to times of order $T \lesssim (\Delta \bar{\Delta})^{-1/2}$. The linear in $T$ behaviour is the manifestation, from the perspective of the cosmological horizon, of the often discussed logarithmic behaviour \cite{ford} in conformal time $\eta \equiv -e^{-T}$ displayed by the massless scalar field two-point function. 
%
%
%
%
\newline\newline
{\textbf{Spacelike separation.}}
Next, we can consider spacelike separated points. For simplicity we will place the fields at the North and South pole of $\Omega_3$, at a fixed global time $T$. This is displayed by the two stars connected by a horizontal dashed line in figure \ref{fig:penrose}. The points are separated by their respective cosmological horizon. In this case, the two-point function is given by
\begin{equation}\label{2ptii}
\langle E | \Phi(T,\boldsymbol{\theta}_N) \Phi(T,\boldsymbol{\theta}_S) | E \rangle \approx \begin{cases}
\alpha (\Delta) e^{-2 \Delta T}  + \alpha (\bar{\Delta}) e^{-2 \bar{\Delta} T} ~, &\text{for $T \gg 1$ and $\Delta\bar{\Delta}>0$~,}\\ \noalign{\vskip4pt}
\frac{(\Delta \bar{\Delta})^{1/4}}{2 \sqrt{2} \pi ^{3/2}} \frac{e^{- \pi \sqrt{\Delta \bar{\Delta}}} }{\sinh ^{{3}/{2}}2 T} \sin \left(2 T \sqrt{\Delta \bar{\Delta}}- \pi/4\right) \,, & \text{for $\Delta \bar{\Delta} \gg 1$ \,, }
\end{cases} 
\end{equation}
where again $\alpha(\Delta)$ is given by (\ref{alpha}). 
The  second line in this expression is not valid at arbitrarily short times; in fact, the approximation breaks down at times $T \approx (\Delta \bar{\Delta})^{-1/2}$ \cite{hyper}. 
It is straightforward to evaluate (\ref{2pt})  at $T=0$, with the spatial points at the North and South poles of $S^3$, to find
\begin{equation}\label{timelike}
\langle E | \Phi(0, \boldsymbol{\theta}_N) \Phi(0, \boldsymbol{\theta}_S) | E \rangle = \frac{\Gamma(\Delta) \Gamma(\bar{\Delta})}{16\pi^2} \approx \frac{ \Delta \bar{\Delta}}{8 \pi } e^{- \pi \sqrt{\Delta \bar{\Delta}}} \,,
\end{equation}
which is finite at finite $\Delta$. The last expression is obtained in the large $\Delta \bar{\Delta}$ limit. 
At finite $T$  and $\Delta \bar{\Delta} >9/4$, the correlator (\ref{2ptii}) (which is real) exhibits a damped oscillatory behaviour, but now with twice the frequency as compared to (\ref{2pti}). This is shown in figure \ref{fig:2pt_space}. Based on the worldline formalism of quantum field theory, in the large mass limit one might expect that the correlator (\ref{timelike}) is obtained by computing the length of a spacelike geodesic connecting the two points. However, there are no real, finite-length geodesics between the two points at $T>0$ \cite{Chapman:2021eyy,proceedings}.  Instead, the oscillatory form of the correlator in the large mass limit $\Delta\bar{\Delta} \gg 1$ -- see the second line in eqs. (\ref{2pti}) and (\ref{2ptii}) -- suggests the existence of complex geodesics that might account for this two-point function. This is indeed the case in dS$_2$ where one can obtain the complex geodesics by analytically continuing the two real geodesics connecting two points on a two-sphere.

For small $\Delta$, the spacelike correlator decreases linearly with time with twice the slope of the timelike case (\ref{timelike2pt}), 
\begin{equation}\label{growing}
\langle E | \Phi(T,\boldsymbol{\theta}_N) \Phi(T,\boldsymbol{\theta}_S) | E \rangle \approx \frac{3}{8 \pi ^2 \Delta \bar{\Delta}}-\frac{T}{4 \pi ^2} \,, \qquad  \text{for $\Delta \bar{\Delta} \ll 1$ and $T \gg 1$} \,, 
\end{equation}
again, up to times of order $T \lesssim (\Delta \bar{\Delta})^{-1/2}$.
%
%
\allowdisplaybreaks
\begin{figure}[H]
        \begin{center}
         \subfigure[Timelike separation]{
                \includegraphics[height=5.65cm]{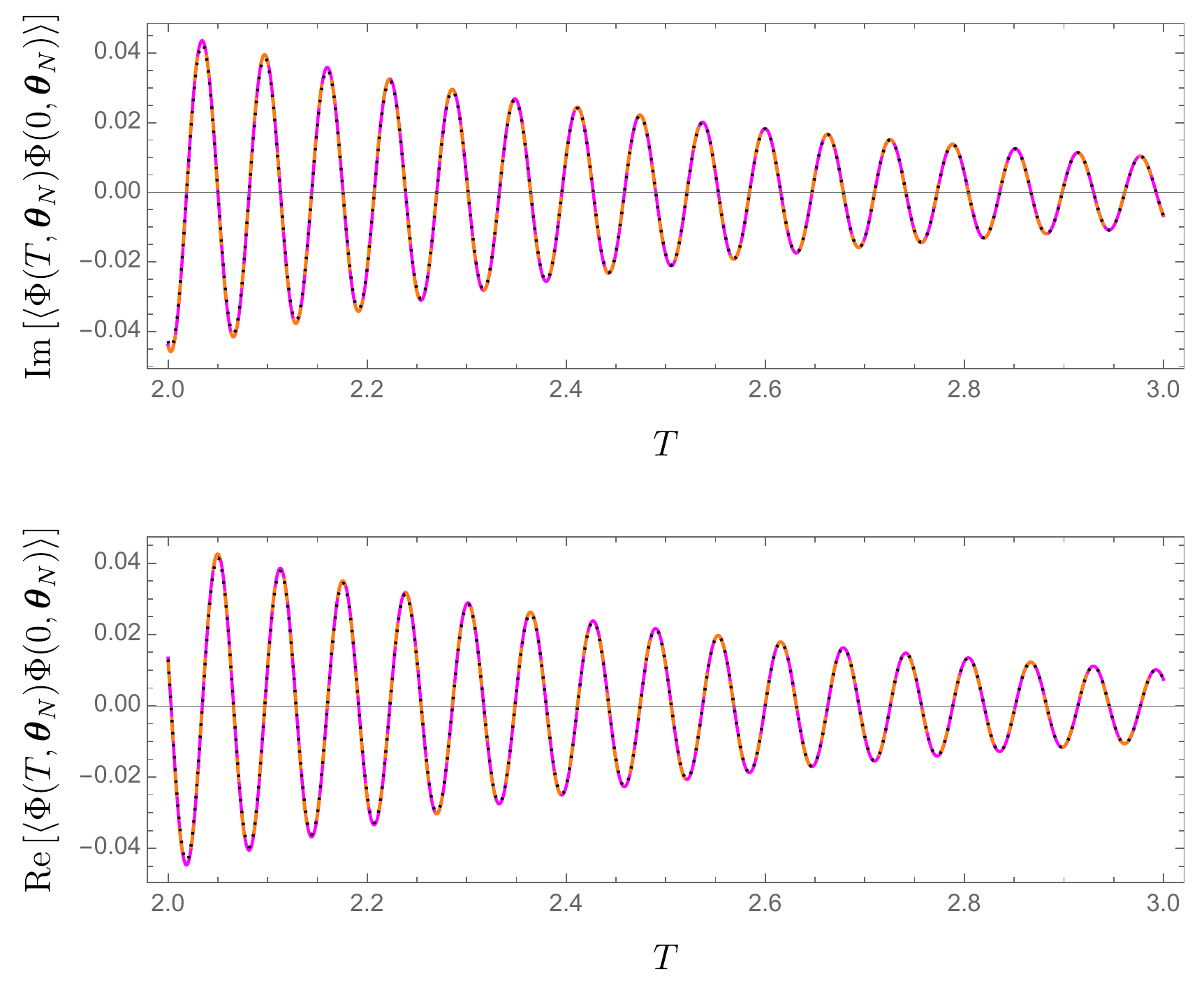}\label{fig:2pt_time}}  \quad\quad
        \subfigure[Spacelike separation]{
                \includegraphics[height=5.2cm]{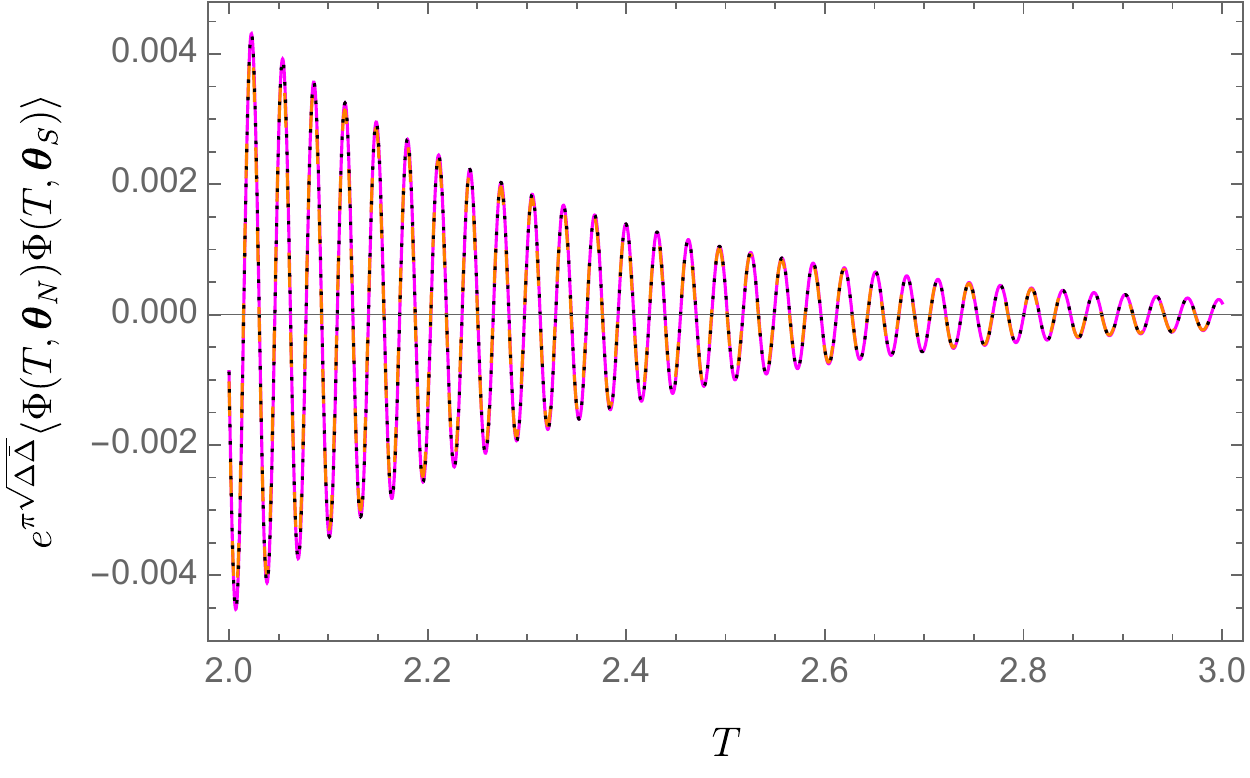} \label{fig:2pt_space}} 
                \end{center}
               \caption{The two-point function of a massive scalar field in dS$_4$ as a function of global time $T$. The mass is fixed to $\sqrt{\Delta \bar{\Delta}} = 100$. The exact two-point function (\ref{2pt}) is plotted in magenta; the large-mass approximation, in dashed orange; and the long-time approximation, in dotted black. The three curves overlap to very good approximation in the time interval plotted both for timelike and spacelike separated points.} 
\end{figure}

\noindent
{\textbf{Interacting theories.}} So far, we have discussed free fields. We now comment briefly on the structure of the two-point function for interacting fields. This follows from a K\"all\'en-Lehmann spectral representation \cite{Bros:1990cu}
\begin{equation}\label{interacting2pt}
\langle E | \tilde{\Phi}(x') \tilde{\Phi}(x) | E \rangle = \int_{\mathcal{C}} \frac{\dd \Delta}{2\pi i} \,  \rho(\Delta) \, \frac{\Gamma(\Delta) \Gamma(\bar{\Delta})}{16\pi^2} \, _2F_1 \left( \Delta,\bar{\Delta}; 2; \frac{1+P_{x',x}}{2} \right)~,
\end{equation}
where $\tilde{\Phi}(x)$ is a scalar operator in some interacting theory.\footnote{Parenthetically, it is also worth noting that if one has (\ref{interacting2pt}) as a function of the coupling $\mu^2$ of the $\Phi^2$ operator, then the coincident point limit encodes the $\mu^2$ derivative of the sphere path integral of the interacting theory.} The integration contour $\mathcal{C}$ is over the principal series $\Delta = 3/2+i\nu$ with $\nu \in \mathbb{R}$ and the spectral function $\rho(\Delta)$ is a non-negative function along $\mathcal{C}$. The free theory corresponds to $\rho(\Delta) = \delta(3/2+ i \nu - \Delta)$. To include the complementary series states we must include the region $i \nu \in (0,3/2)$ in the contour.\footnote{It should be noted here that the complementary series obey some unusual properties. For instance, they are not contained in the tensor product of any two non-complementary series irreducible representations of $SO(4,1)$ and moreover only contribute in an isolated way to tensor products among themselves \cite{Dobrev:1976vr}.} More generally, if we take $\rho(\Delta)$ to be a meromorphic function with poles at $\text{Re} \, \Delta > 3/2$, as assumed in \cite{Sleight:2020obc,DiPietro:2021sjt,Hogervorst:2021uvp},  we can close the contour in (\ref{interacting2pt}). The timelike separated two-point function in the interacting theory will now decay by an amount governed by the pole $\Delta_*$ of $\rho(\Delta)$ with the smallest real part. Provided $\Delta_*$ has an imaginary value it will continue to exhibit oscillations. 
\newline
\begin{center} {\textit{Holographic considerations}} \end{center}

From a holographic perspective, we can  contrast the dynamical features of the cosmological horizon to large black holes in an asymptotically anti-de Sitter world which have come to be viewed as strongly coupled, chaotic, dissipative states. For such black holes, correlation functions of heavy fields such as (\ref{2pt}) computed in the black hole background decay exponentially rather than in an oscillatory fashion, with an exponent whose real part grows with the mass \cite{Birmingham:2001pj}. The decay is meant to capture, holographically, the dissipative nature of the thermal state dual to the black hole horizon. The oscillatory behaviour of (\ref{2pt}) in the static patch, and more markedly the non-decaying correlations of light particles (\ref{growing}) are in sharp contrast to the black hole. (From a global perspective infrared effects such as (\ref{growing}) have been shown \cite{Anninos:2011kh} to enjoy features in common with glassy systems.) This may put to question the idea that the microstates of the de Sitter horizon are thermalised in a standard fashion. 

It is also worth comparing the phenomena to BPS black holes. Due to a gap in the spectrum of low lying excitations, BPS black holes exhibit oscillatory correlations at late times  \cite{Turiaci}. Moreover, the microstates of BPS black holes have exactly degenerate energies that do not mix -- it is unclear whether they can thermalise in an ordinary way and whether one can apply the standard assumptions of ergodicity to the set of BPS configurations. 
{\subsection{Dynamical probes on a dynamical geometry}}\noindent
The oscillations observed in (\ref{2pti}) and (\ref{2ptii}) are a pertinent qualitative feature of the cosmological horizon. They are a  physical manifestation of the fact that heavy particles are increasingly shielded from the cosmological horizon.  Once the geometry becomes dynamical, however, correlation functions such as (\ref{2pti}) and (\ref{2ptii}) suffer from  not being diffeomorphism invariant. We turn now to non-linear dynamical features.
\newline\newline
\textbf{Gao-Wald theorem as a non-linear feature.}
The Gao-Wald theorem \cite{Gao_Wald} states that for matter obeying the null energy condition and null generic energy condition, a null geodesically complete and globally hyperbolic spacetime with compact Cauchy surface $\Sigma$ cannot exhibit a particle horizon. In particular, at sufficiently late times the past light cone of a spacetime point in a slightly perturbed de Sitter spacetime intersects all of $\Sigma$. The requirement of a slightly perturbed de Sitter spacetime is to ensure that the null generic energy condition is also satisfied. If we now arrange the small perturbation\footnote{By small perturbation we mean that the perturbed spacetime retains a smooth metric that has both an asymptotic past infinity $\mathcal{I}^-$ and future infinity $\mathcal{I}^+$. Such perturbations were proven to exist at the non-linear level in general relativity in a theorem of Friedrich \cite{friedrich}.} to be a localised but smooth source of matter encoded in a small spacelike neighbourhood $\mathcal{N}$ (smaller than the de Sitter scale) near $\mathcal{I}^-$, the resulting spacetime will be such that the past lightcone at a point lying strictly before $\mathcal{I}^+$ intersects the full Cauchy surface $\Sigma$. Such a configuration can be compared to the pure de Sitter case where only the point at $\mathcal{I}^+$ will intersect $\Sigma$. In the perturbed spacetime a null ray emitted from the antipodal point of the Cauchy surface near $\mathcal{I}^-$ will fall inside the past light cone of a spacetime point $\mathcal{P}$ connected to $\mathcal{N}$ by a timelike curve. 

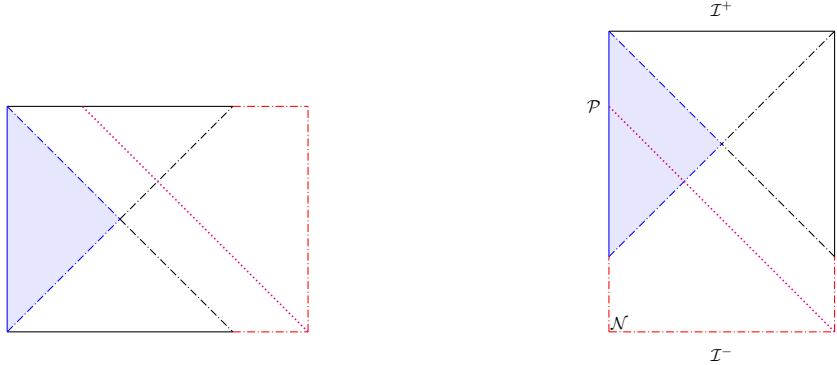
\begin{figure}[H]
\begin{center}
\begin{tikzpicture}[scale=2.]

 \draw[blue] (4,.5) -- (4,2); 
  \draw[red, densely dashdotted] (4,0) -- (4,.5); 
   \draw[red, densely dashdotted] (5.5,0) -- (5.5,.5);  
      \draw[red, densely dashdotted] (4,0) -- (5.5,0);  
 \draw[densely dashdotted] (5.5,.5) -- (4.75,1.25);  
 \draw[name path=A, densely dashdotted, blue] (4,2) -- (4.75,1.25);    
  \draw[densely dashdotted] (5.5,2) -- (4.75,1.25);  
\draw[name path=B, densely dashdotted, blue] (4,.5) -- (4.75,1.25);    
    \node[scale=.6] at (4.75, -.15)   {$\mathcal{I}^-$ ~ };
     \node[scale=.6] at (4.75, 2.15)   {$\mathcal{I}^+$ ~ };    
  \node[scale=.6] at (4.07,.06)   {$\mathcal{N}$ ~ };         
      \draw (5.5,2) -- (5.5,.5);   
    \draw (4,2) -- (5.5,2);     
  \draw (1.5,0)--(0,0);    
  \draw (1.5,1.5) -- (0,1.5);
    \draw[red, densely dashdotted] (1.5,0)--(2,0);      
    \draw[red, densely dashdotted] (1.5,1.5)--(2,1.5);    
    \draw[red, densely dashdotted] (2,0)--(2,1.5);    
    \draw[name path=C,  blue] (0,0)--(0,1.5);  
  \draw[name path=D, densely dashdotted, blue] (0,0) -- (.75,.75);     
     \draw[densely dashdotted]  (.75,.75)--(1.5,1.5);     
   \draw[name path=F, densely dashdotted, blue] (0,1.5) -- (.75,.75);      
    \draw[densely dashdotted] (.75,.75) -- (1.5,0);   
     \node[scale=.6] at (3.9,1.5)   {$\mathcal{P}$ ~ };       
\draw[densely dotted, magenta,semithick] (2,0) -- (.5,1.5);     
\draw[densely dotted, magenta,semithick] (5.5,0) -- (4,1.5);
    \tikzfillbetween[of=A and B]{blue, opacity=0.1};
    \tikzfillbetween[of=D and F]{blue, opacity=0.1};    
\end{tikzpicture}
\end{center}
\caption{\small{Gao-Wald effect on a large AdS black hole (left) and the cosmological horizon (right). We denote by the dotted line an incoming light-ray.}}
\label{fig:sewing}
\end{figure}
\noindent
This is one of the most notable classical features exhibited by the cosmological horizon at the non-linear level. It can be contrasted  to the case of black holes absorbing energy pulses. For large black holes in AdS, as explored in  \cite{butterfly_SS}, previously accessible regions of space become causally inaccessible. The Gao-Wald effect suggests that the cosmological horizon responds in an unusual way to localised disturbances as compared to black hole horizons. 
\newline
\begin{center} {\textit{Further holographic considerations}} \end{center}
We can contrast the reaction of perturbed large AdS black holes and the cosmological horizon. Holographically, the reaction of large AdS black holes to shockwaves is meant to capture the disruption of the delicately tuned entanglement structure of the thermofield double state \cite{butterfly_SS}. For a black hole to mimic the Gao-Wald behaviour observed for de Sitter one would have to send  into the black hole horizon energy that violates the null energy condition. In AdS/CFT this can be achieved  \cite{Gao:2016bin}, for instance, by coupling the operators of a  pair of entangled conformal field theories in the thermofield double state dual to the two-sided AdS black hole. Moreover, the spacelike separated correlations (\ref{2ptii}) exhibit oscillations as global time evolves, behaving again in stark contrast \cite{Chapman:2021eyy,Shaghoulian:2022fop} to the analogous correlations of the thermodouble state dual to large AdS black holes which exhibit exponential decay. 

Dynamically, then, it is unclear that the de Sitter horizon will have a microphysical description that mimics, even qualitatively, that of a large AdS black hole -- namely, {that of} a chaotic, dissipative, and strongly coupled  liquid state. The oscillatory nature of the correlations, both across the cosmological horizon and along the timelike worldline, as well as the opening up of spacetime due to the Gao-Wald effect are suggestive of a microphysical system distinct to the two-sided large anti-de Sitter black hole. 
\newline\newline
\textbf{AdS$_2 \times S^2$ $\to$  dS$_4$ flows.} To make the comparison to large AdS black holes sharp, and gain some insight toward a microphysical picture, it is convenient to introduce a solid worldtube $\mathcal{W}$ (of topology $\mathbb{R} \times B^3$) whose boundary $\partial \mathcal{W}$ (of topology $\mathbb{R} \times S^2$)  is subject to a specified set of boundary conditions. {Such a setup has been considered in various works \cite{Friedrich:1998xt,An:2021fcq, Witten:2018lgb}, where it has been noted that the standard Dirichlet problem does not produce a well-posed boundary value problem. We elaborate on  this  in appendix \ref{IBVP}, where we also discuss a modified boundary value problem fixing the conformal class of the metric on a timelike boundary as conjectured in \cite{An:2021fcq}.} 

Here, instead, consider the possibility of embedding the static patch within a geometry with an asymptotic timelike boundary at spatial infinity \cite{Centaur1,Centaur2,Centaur3}.\footnote{Considerations along a slightly different path involving the AdS$_3$/CFT$_2$ correspondence are discussed in \cite{Coleman:2021nor}.} In particular, we study a deformed metric that  resembles an  AdS$_2\times S^2$ type geometry near $\partial \mathcal{W}$. As we now show, such geometries can be made compatible with the null energy condition. Assume, for simplicity, a static, spherically symmetric spacetime with metric given by
\begin{equation}\label{flow_geometry}
ds^2 = -f(\rho)\dd t^2 + \frac{\dd\rho^2}{f(\rho)}+ g(\rho)^2 \left( \dd \theta^2 + \sin^2 \theta \dd \varphi^2 \right)~.
\end{equation}
The functions $f(\rho)$ and $g(\rho)$ are constrained by the null energy condition, as shown in (\ref{eq:inequalities}) of appendix \ref{app:flow}, where more details are described. 
The following example satisfies the null energy condition for a given range of $\rho$ and admits a macroscopic piece of dS$_4$ in the interior of the spacetime
\begin{equation}\label{egflow}
f(\rho) = \frac{2\rho \ell - \rho^2}{\ell^2} + \left( \left( \frac{\rho - \rho_c}{\la} \right)^2 + 1 \right) \left(  \frac{1 + \tanh \frac{\rho-\rho_c}{\varepsilon}}{2}\right) \,, ~\quad g(\rho) = \ell\left(1-\frac{\rho}{\ell}\right) \,.
%
\end{equation}
For $\ell_{\scaleto{\text{AdS}}{4.5pt}} \ll \ell$, the geometry (\ref{egflow}) interpolates between the static patch of dS$_4$ in the region $0 \leq \rho \lesssim \rho_c$, to an approximately AdS$_2$ in the region $\rho_c \lesssim \rho \leq \ell$. The width of the interpolating region is controlled by the parameter $\varepsilon$, which is chosen to be of order $\ell_{\scaleto{\text{AdS}}{4.5pt}}$, see figure \ref{fig:flowf}.

\begin{figure}[H]
	\begin{center}
	\includegraphics[scale=0.36]{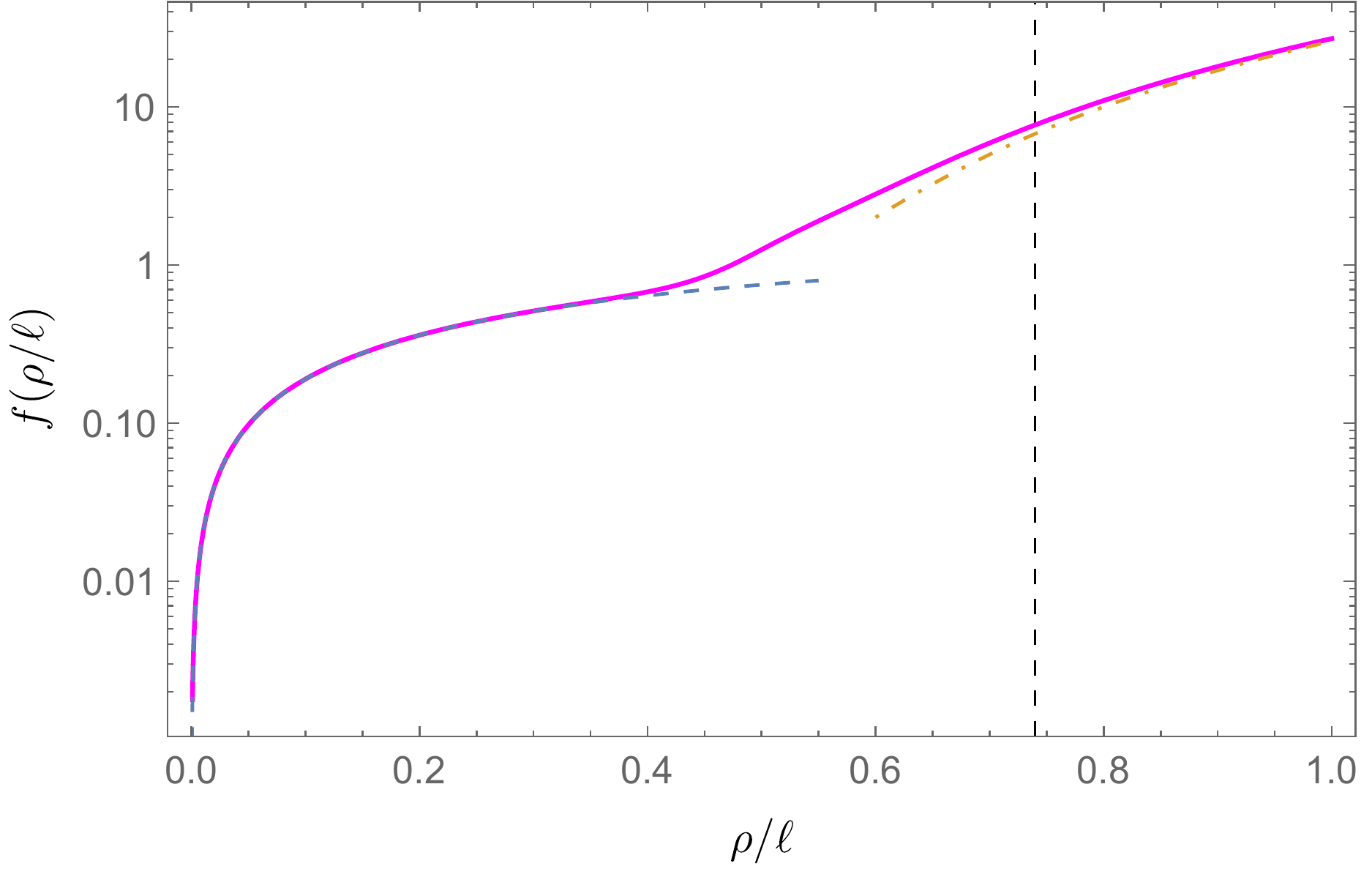}
	\caption{The function $f$ in (\ref{egflow}), with $\ell_{\scaleto{\text{AdS}}{4.5pt}}/\ell = 0.1$, $\rho_c = 0.5 \ell$ and $\varepsilon = 0.05 \ell$. The vertical axis is on a logarithmic scale. For reference, we also show the de Sitter type behaviour in dashed blue, and the AdS$_2$ type behaviour in dot-dashed yellow. The vertical dashed line indicates the value of $\rho/\ell$ for which the null energy condition is saturated for this choice of parameters.}
	\label{fig:flowf}
	\end{center}
\end{figure}
\noindent
The null energy condition is only satisfied up to a certain value of the radial coordinate $\rho_b \lesssim \ell$. The proper distance between the horizon and the boundary grows logarithmically with $\rho_b/\ell_{\scaleto{\text{AdS}}{4.5pt}}$ and is large when $\rho_b \sim \ell$ and $\ell_{\scaleto{\text{AdS}}{4.5pt}} \ll \ell$.

In the absence of a theory for which (\ref{egflow}) is a solution it is difficult to propose and test a set of boundary conditions for a more general class of metrics, and the stability of the spacetime. We simply note that the following boundary conditions
\begin{equation}\nonumber
\begin{bmatrix}
   g_{tt}   & 0 & g_{t\theta}  & g_{t\varphi} \\
    0 & g_{\rho\rho}  & 0 & 0 \\
   g_{t\theta}  & 0 &  g_{\theta\theta} &  g_{\theta\varphi}   \\
    g_{t\varphi} & 0  &  g_{\theta\varphi}   &  g_{\varphi\varphi} 
\end{bmatrix} = 
\begin{bmatrix}
    \rho^2 + \mathcal{O}(\rho) & 0 &  \mathcal{O}(\rho^{-1}) & \mathcal{O}(\rho^{-1})  \\
    0 & \rho^{-2} + \mathcal{O}(\rho^{-3}) & 0 &   0 \\
    \mathcal{O}(\rho^{-1}) & 0 & \ell^2 -2 \ell \rho+ \mathcal{O}\left(\rho^{-1}\right)   & \mathcal{O}(\rho^{-1})   \\
     \mathcal{O}(\rho^{-1}) & 0& \mathcal{O}(\rho^{-1}) & \ell^2 -2 \ell \rho+ \mathcal{O}\left(\rho^{-1}\right)  
\end{bmatrix}~,
\end{equation}
are compatible with (\ref{flow_geometry}) and (\ref{egflow}) near the AdS$_2 \times S^2$ boundary. We have chosen the gauge $g_{\mu\rho} = 0$, selected the units $\ell_{\scaleto{\text{AdS}}{4.5pt}} = 1$, and expanded in $\rho \gg 1$ while keeping $\rho/\ell \lesssim 1$. The ultimate fate of the above or any boundary conditions rests on the existence of a theory permitting a class of such asymptotic configurations and a well-defined, finite set of boundary charges.   It would also be interesting to study this from the perspective of a dimensionally reduced, two-dimensional model, using techniques such as those developed in \cite{Grumiller:2021cwg}.


From the worldtube perspective of the AdS$_2$ boundary we can now arrange for a boundary value problem whereby the matter is localised at a spacelike surface on $\partial \mathcal{W}$ at some early time.  It is subsequently propelled into the spacetime, potentially in the form of a shockwave.\footnote{The simplest analytically known shockwave solution in de Sitter \cite{Sfetsos:1994xa} takes a rather different form, namely it consists of a spherically symmetric pulse that is located on the cosmological horizon throughout its evolution. This solution does not lend itself to the setup we have in mind. It would be interesting to construct shockwave solutions that are closer to the case at hand.} The response is then measured at some later time with respect to the coordinate system endowed on $\partial \mathcal{W}$. In the presence of $\mathcal{W}$, the spacetime has a timelike boundary and is no longer globally hyperbolic, so care must be taken in applying the Gao-Wald theorem. The two-dimensional version   of this setup, whereby an asymptotically AdS$_2$ space flows to a dS$_2$ interior, has been shown \cite{Centaur2} to exhibit a phenomenon akin to that of Gao and Wald.  It would be interesting to generalise the Gao-Wald theorem to the current setup.  
\newline
\begin{center} {\textit{Finiteness $\&$ AdS$_2$}} \end{center}

Regarding the nature of degrees of freedom, the deformed geometries (\ref{egflow}) endowed with an AdS$_2$ boundary are suggestive of a quantum mechanical dual description. Being quantum mechanical, the putative duals are built from a finite number of degrees of freedom. For the cases considered thus far, the cutoff near the AdS$_2$ boundary of (\ref{flow_geometry}) cannot be pushed arbitrarily far since we require that the size of the two-sphere is large in Planck units (such that $g(\rho)^2/G_4 \gg 1$) to maintain semiclassical control. Interestingly, it seems that generically the null energy condition does not allow the size of the two-sphere to become too small. 
In AdS/CFT, a cutoff near the boundary is interpreted as an ultraviolet cutoff in the dual CFT. When the CFT is a quantum mechanical theory with a spectrum bounded from below, imposing an ultraviolet cutoff leads to a finite number of states. 
In the spirit of various discussions relating AdS/CFT to condensed matter systems, it would be interesting to identify a simple toy model of a many-body quantum system that exhibits some of the de Sitter like features we have described (a dS/CMT correspondence \cite{Centaur2} of sorts), or establish that they are incompatible with any such interpretation. 

It is striking to note that the size $g(\rho)$ of the celestial sphere decreases in size as we radially approach the AdS$_2$ boundary. It is unclear how such a feature might be accommodated in the dual theory which in all known cases describes a sphere of increasing size toward the AdS boundary. {Perhaps, again, unusual features such as `negative' microphysical degrees of freedom might be necessary to accommodate a celestial sphere that decreases in size. This may also tie into the Gao-Wald behaviour of the cosmological horizon, as compared to that of black hole horizons, upon absorbing a pulse of energy.} 

\section{Euclidean signature and localisation}\label{sec3}

Localisation of a path integral manifests itself in a significant reduction of the degrees of freedom to be path integrated over, and may tie to broader considerations necessitating an effective reduction in degrees of freedom in theories of quantum gravity. 

In this section, we speculate whether the Euclidean manifestation of finiteness in quantum de Sitter is that the underlying gravitational path integral localises.  For the sake of concreteness, we focus on theories of  two-dimensional quantum gravity admitting a  two-dimensional Euclidean de Sitter saddle, and in particular propose a two-dimensional theory of supergravity that exhibits supersymmetric localisation.
\newline\newline
\textbf{BRST localisation.} In the Weyl gauge two-dimensional quantum gravity can be expressed in terms of a path integral weighted by the  Liouville action.  Unless otherwise specified, the matter theory will be a two-dimensional CFT with central charge $c_{\text{m}} > 25$ that we minimally couple to gravity (see appendix \ref{Liouville_primer} for further details). 
The two-sphere path integral (\ref{Zh}) is given by
\begin{equation}\label{Zgrav2D_weyl2}
\mathcal{Z}_{S^2} = e^{2\vartheta}\int\frac{[\mathcal{D}\varphi][\mathcal{D}\Phi][\mathcal{D}\mathfrak{b}\mathfrak{c}]}{\text{vol}_{PSL(2,\mathbb{C})}}\,e^{- S}~,
\end{equation}
where $S\equiv S_{tL}[\varphi]+S_{\text{CFT}}[\Phi] + S_{\text{gh}}[\mathfrak{b},\mathfrak{c}]$ combines the action of timelike Liouville theory, matter, and ghost sector. 
The timelike Liouville action is given by
\begin{equation}\label{tL_sec3}
 S_{tL}[\varphi] = \frac{1}{4\pi} \int_{S^2} \dd^2 x \sqrt{\tilde{g}}\left(-\tilde{g}^{ij}\partial_i \varphi \partial_j\varphi -q \tilde{R}\varphi + 4\pi \Lambda e^{2\beta\varphi}\right)~,
\end{equation}
where $\varphi$ is the Weyl factor of the physical metric  $g_{ij}= e^{2\beta \varphi}\tilde{g}_{ij}$, $q= \beta^{-1}-\beta$, and $\tilde{R}$ is the Ricci scalar of the background metric $\tilde{g}_{ij}$, which we take to be the standard round two-sphere. Vanishing of the conformal anomaly imposes the condition $c_{tL}+ c_{\text{m}}-26=0$, where $c_{tL}= 1-6q^2$ and $c_{\text{gh}}=-26$ is the $\mathfrak{b}\mathfrak{c}$-ghost system central charge. The path integral (\ref{Zgrav2D_weyl2}) admits a  two-dimensional Euclidean de Sitter saddle and a non-trivial semiclassical $\beta \rightarrow 0^+$ loop expansion \cite{timelike, Muhlmann:2022duj}. 

The BRST symmetry is generated by a nilpotent BRST charge $\mathcal{Q}_B$.
The area operator $\mathcal{V}_\beta \equiv \mathfrak{\tilde{c}}\mathfrak{c}\, e^{2\beta\varphi}$ in (\ref{Zgrav2D_weyl2}) has conformal dimension $\Delta_\beta=\bar{\Delta}_\beta=1$ and is hence a physical operator. It was noted in \cite{Polchinski_WI} that the following local operator relation is satisfied 
\begin{equation}\label{WI}
\mathcal{Q}_B \mathcal{V} =\beta \Lambda   \mathcal{V}_\beta + \mathcal{K}~,\quad\quad  \mathcal{V} \equiv -\frac{1}{4\pi}\times L_{-1}^{\varphi}\mathfrak{\tilde{b}}_{-1} \mathfrak{\tilde{c}}\mathfrak{c}\,\varphi
\end{equation}
where $L_{-1}^\varphi$ is the Liouville Virasoro generator and $\mathcal{K}$ is an expression asymmetric in the ghost number. 
A $\Lambda$-derivative of the path integral (\ref{Zgrav2D_weyl2}) leads to
\begin{equation}\label{Z_ind_Lambda}
\beta\Lambda\frac{\dd}{\dd\Lambda}\mathcal{Z}_{S^2} \propto -e^{2\vartheta}\int {[\mathcal{D}\varphi][\mathcal{D}\Phi][\mathcal{D}\mathfrak{b}\mathfrak{c}]} e^{- S}\mathcal{Q}_B \mathcal{V}=-e^{2\vartheta}\int {[\mathcal{D}\varphi][\mathcal{D}\Phi][\mathcal{D}\mathfrak{b}\mathfrak{c}]} \mathcal{Q}_B  \left( e^{-S}\mathcal{V} \right)~,
\end{equation}
where the proportionality constant is $\Lambda$-independent. In the absence of boundary or curvature effects and under the assumption that the  measure is $\mathcal{Q}_B $-invariant, the final expression in (\ref{Z_ind_Lambda}) vanishes. This, in turn, would  predict that the non-analytic behaviour in $\Lambda$ for the two-sphere partition function vanishes. However, arguments based on semiclassical considerations of the  Liouville path-integral \cite{timelike,area,Zamolodchikov:1982vx},  the DOZZ formula for (spacelike) Liouville theory \cite{DO, ZZ}, and the representation of  two-dimensional quantum gravity as a matrix model \cite{QG_MI1, QG_MI2,QG_MI3} indicate that $\mathcal{Z}_{S^2}$ depends non-analytically on $\Lambda$.

It follows that $\mathcal{Z}_{S^2}$ must localise onto curvature and boundary effects  \cite{Polchinski_WI}. Boundary effects  arise in correlation functions of $n > 2$ operators in the regime where the operators approach each other. The round $S^2$ metric can be mapped onto the flat complex plane with a collection of curvature singularities where we insert the $n$ vertex operators. This leads to curvature contributions in the expectation values also. For genus zero,  the expectation value of $n = 4$ area operators $\mathcal{V}_\beta$ is related to the expectation value of $n=3$ area operators as follows  \cite{Polchinski_WI}
\begin{equation} \nonumber
 \Lambda  \left\langle   \int \dd^2 z \, e^{2\beta\varphi(z,\bar{z})} \, \mathcal{V}_\beta(z_1,\bar{z}_1)\mathcal{V}_\beta(z_2,\bar{z}_2)\mathcal{V}_\beta(z_3,\bar{z}_3)\right\rangle  \\ \nonumber
= \left(\frac{q}{\beta}-3\right)\big\langle \mathcal{V}_\beta(z_1,\bar{z}_1)\mathcal{V}_\beta(z_2,\bar{z}_2)\mathcal{V}_\beta(z_3,\bar{z}_3)\big\rangle~,
\end{equation}
where one has used that expectation values involving $\mathcal{K}$ vanish. 
Taking this into account we obtain the equation
\begin{equation}
\Lambda\frac{\dd^4 \mathcal{Z}_{S^2}}{\dd\Lambda^4}= \left(\frac{q}{\beta}-3\right)\frac{\dd^3 \mathcal{Z}_{S^2}}{\dd\Lambda^3}~.
\end{equation}
Consequently we infer $\mathcal{Z}_{S^2} \propto \Lambda^{q/\beta}$. 

When the matter CFT in (\ref{Zh}) is one of the  minimal models, the relation (\ref{WI}) can be extended to an infinite set of relations involving additional ghost insertions. 
The gravity partition function in these cases is related to the KdV hierarchy leading to an infinite set of linear constraints \cite{VVD} on the square root of the partition function. It is argued in \cite{Polchinski_WI} that the infinite number of operator relations are strong enough to localise the gravitational path integral onto an ordinary matrix integral \cite{QG_MI1, QG_MI2, QG_MI3}. (Perhaps a parallel can be drawn to the infinite set of null states discussed in \cite{marolf,Anous:2020lka}.) Using the operators in \cite{Polchinski_WI}, the authors of \cite{Zamolodchikov:2003yb,Belavin:2006ex}  introduced the so called higher equations of motion for Liouville theory. These reduce the calculation of correlation numbers -- which are correlation functions integrated over the entire surface -- to boundary and curvature terms. 
\newline\newline
\textbf{Supersymmetric localisation.} We now turn to two-dimensional $\mathcal{N}=2$ supergravity, as a somewhat richer example of a theory of two-dimensional gravity that localises on the two-sphere. The $\mathcal{N}=2$ supergravity multiplet is given by the graviton $g_{ij}$, the Dirac gravitino $\chi_i$, and a $U(1)$ gauge field $A_i$. In a supersymmetric extension of the Weyl gauge, $\mathcal{N}=2$ supergravity can be expressed as $\mathcal{N}=2$  super Liouville theory \cite{Ivanov_Krivonos, Antoniadis:1990mx, HK,Murthy:2013mya}. For simplicity we focus on the case with vanishing $U(1)$ background flux and $S^2$ topology. In Euclidean signature $\mathcal{N}=2$  super Liouville theory is built from a chiral mulitplet $(\varphi, \psi, F)$  and an anti-chiral multiplet $(\widetilde{\varphi},\widetilde{\psi},\widetilde{F})$. It is a superconformal field theory of central charge $c_{sL} = 3+6/b^2$. The semiclassical limit is given by taking $b \to 0^+$. In this limit, the classical saddle on an $S^2$ topology corresponds to a physical metric which is complex, much like what happens for non-supersymmetric spacelike Liouville theory. 

To obtain a real Euclidean de Sitter saddle, we consider a timelike version of $\mathcal{N}=2$ super Liouville theory. The resulting Lagrangian on a two-sphere topology is 
\begin{multline}\label{SN2sl}
\mathcal{L}_{\text{tsL}} = - \tilde{g}^{ij}\partial_i\varphi \partial_j \widetilde{\varphi} + i \btpsi \snabla \psi + \widetilde{F}F - \frac{1}{2\beta}\tilde{R}(\varphi + \widetilde{\varphi})\cr
+ i\left(F \frac{\partial W}{\partial\varphi} -\frac{1}{2}\frac{\partial^2 W}{\partial\varphi^2} \overline{\psi}\psi  \right)+i\left(\widetilde{F}\frac{\partial \widetilde{W}}{\partial\tvarphi} +\frac{1}{2}\frac{\partial^2 \widetilde{W}}{\partial\tvarphi^2} \btpsi \tpsi\right) ~,
\end{multline}
where $W= \mu e^{\beta\varphi}$ and $\widetilde{W} = \mu^* e^{\beta\widetilde{\varphi}}$, $\mu\in \mathbb{C}$, and $\tilde{R}$ denotes the Ricci scalar of the background metric $\tilde{g}_{ij}$. The conjugate for a spinor is defined by $\overline{\lambda} = \lambda^T C$ where $C$ denotes the charge conjugation matrix, and the Euclidean $\gamma$-matrices obey $\gamma_i^T = - C\gamma_i C^{-1}$. It is natural to postulate that timelike super-Liouville theory is a (non-unitary) superconformal field theory of central charge $c_{\text{stL}} = 3- 6/\beta^2$. Evidence for this can be provided by computing the corresponding Euclidean path integral
\begin{equation}\label{Z_SUSY}
\mathcal{Z}_{\scaleto{{S^2}}{5.5pt}}^{\scaleto{{\mathcal{N}=2}}{4pt}}= \int \frac{[\mathcal{D}\Phi][\mathcal{D}\widetilde{\Phi}]}{{\text{vol}_{OSP(2|2,\mathbb{C})}}}\,e^{-\frac{1}{4\pi} \int_{S^2} \dd^2 x \sqrt{\tilde{g}}\,\mathcal{L}_{\text{tsL}}}~,\quad\quad [\mathcal{D}\Phi] \equiv [\mathcal{D}\varphi][\mathcal{D}\psi][\mathcal{D}F]~,
\end{equation}
where $OSP(2|2,\mathbb{C})$ denotes the residual gauge group on the two-sphere after fixing Weyl gauge. Integrating out the auxiliary $F$-term allows us to read off the cosmological constant $\Lambda = +\beta^2\mu\mu^*/4\pi$ (\ref{tL_sec3}). Consequently, the action (\ref{SN2sl}) admits a Euclidean two-dimensional de Sitter geometry as a saddle point solution of the physical metric $g_{ij} = e^{\beta(\varphi + \tvarphi)} \tilde{g}_{ij}$. The non-unitarity of  timelike super-Liouville theory renders supersymmetry and a Euclidean de Sitter saddle compatible. Assuming a constant saddle, the saddle point solutions for the bosonic fields in (\ref{SN2sl}) are given by
\begin{equation}
 (\varphi+\tvarphi)_*= \frac{1}{\beta}\log\frac{\tilde{R}}{2\mu\mu^* \beta^4}~,\quad F_*= -i\frac{\tilde{R}}{2\mu\beta^3}e^{-\beta\varphi_*}\,,~\quad  \widetilde{F}_*= -i\frac{\tilde{R}}{2\mu^*\beta^3}e^{-\beta\tvarphi_*}~,
\end{equation}
with the constant part of $(\varphi - \tvarphi)$ unfixed. A semiclassical expansion around this saddle leads to a non-vanishing $\Lambda$-dependent $\mathcal{Z}_{\scaleto{{S^2}}{5.5pt}}^{\scaleto{{\mathcal{N}=2}}{4pt}}$. As will be explored in \cite{DPB}, the action associated to the Lagrangian (\ref{SN2sl}) is  exact under $\mathcal{N}=2$ supersymmetry transformations. This would naively imply that $\mathcal{Z}_{\scaleto{{S^2}}{5.5pt}}^{\scaleto{{\mathcal{N}=2}}{4pt}}$ is essentially a constant, and in particular independent of  $\Lambda$ and $c_{tL}$. 
To agree with the semiclassical expansion, much like (\ref{Zgrav2D_weyl2}), the path integral $\mathcal{Z}_{\scaleto{{S^2}}{5.5pt}}^{\scaleto{{\mathcal{N}=2}}{4pt}}$ (\ref{Z_SUSY}) must therefore contain boundary and curvature effects. A careful treatment of timelike super Liouville will be provided in  \cite{DPB}.
\newline\newline
From a broader perspective, and making contact with our previous considerations, it is tempting to interpret the phenomenon of localisation of the gravitational path integral  \cite{Polchinski_WI,Banerjee:2009af,deWit:2018dix,Jeon:2018kec,Dabholkar:2010uh}, which leads to a significant reduction in the degrees of freedom being integrated over, as a (Euclidean) manifestation of the more finite features of quantum de Sitter space \cite{Banks_dS,Erik_dS, Bousso:2000nf, Fischler_talk,Banks:2001yp,Banks:2018ypk,Banks:2020zcr}. In the context of supersymmetry and supegravity, this reduction is often understood as a restriction of the theory to a BPS subspace of microstates \cite{Dabholkar:2010uh,Dabholkar:2011ec,Dabholkar:2014ema}. For de Sitter,  this restricted subspace may be the entire holographic theory.

\section*{Acknowledgements}

It is a pleasure to acknowledge Tarek Anous, Costas Bachas, Lorenz Eberhardt, Pietro Benetti Genolini, Frederik Denef, Diego Hofman, Chawakorn Maneerat, Sameer Murthy, Luigi Tizzano, and David Vegh for useful discussions. 
D.A. is funded by the Royal Society under the grant The Atoms of a deSitter Universe. The work of D.A.G. is funded by a UKRI Stephen Hawking Fellowship. B.M. is supported in part by the Simons Foundation Grant No. 385602 and the Natural Sciences and Engineering Research Council of Canada (NSERC), funding reference number SAPIN/00047-2020.

\appendix

\section{Liouville primer}\label{Liouville_primer}
This appendix provides the necessary details for two-dimensional Liouville theory.  In Euclidean signature the gravitational path integral in two-dimensions on a compact Riemann manifold $\Sigma_h$ of genus $h$ is given by 
\begin{equation}\label{Zgrav2D}
\mathcal{Z}_{\Sigma_h} =  e^{\vartheta \chi_h} \int  \frac{\mathcal{D} g_{ij}}{\text{vol} \, \mathcal{G}_2}  e^{-\Lambda \int_{\Sigma_h} \sqrt{g}} \, Z_{\text{matter}}[g_{ij}]~,
\end{equation}
where we used that the integral over the Ricci scalar in two-dimensions is topological and yields the Euler character $\chi_h= 2-2h$ and $\vartheta$ is a coupling associated to genus. $\Lambda>0$ is the cosmological constant coupled to the area term in the exponent. We further included contributions of a matter CFT. In two-dimensions we can furthermore exploit the two diffeomorphisms to write the metric in terms of a fixed fiducial part $\tilde{g}_{ij}$ and a single degree of freedom appearing in the form of the Weyl factor $\varphi(x)$ such that 
\begin{equation}
g_{ij}= e^{2b\varphi}\tilde{g}_{ij}~.
\end{equation} 
It has been conjectured \cite{David,Distler_Kawai} that the change in variables from $g_{ij}$ to $\varphi$ in (\ref{Zgrav2D}) produces a non-trivial Jacobian, such that
\begin{eqnarray}\label{Zgrav2D_weyl}
\mathcal{Z}_{\text{grav}} &=&e^{\vartheta \chi_h} Z_{\text{CFT}}[\tilde{g}_{ij}] Z_{\text{gh}}[\tilde{g}_{ij}] \int\frac{[\mathcal{D}\varphi]}{\text{vol}\,{\mathcal{G}_2}}e^{- S_L[\varphi]}~,\cr
 S_L[\varphi]&=& \frac{1}{4\pi} \int_{\Sigma_h} \dd x^2 \sqrt{\tilde{g}}\left(\tilde{g}^{ij}\partial_i \varphi \partial_j\varphi + Q \tilde{R}\varphi + 4\pi \Lambda e^{2b\varphi}\right)~,
\end{eqnarray}
where $ S_L[\varphi]$ is known as the Liouville action. In the above expressions $Q= b+b^{-1}$ and $c_L= 1+6Q^2$. Liouville theory is a two-dimensional conformal field theory. Vertex operators are given by $\mathcal{V}_\alpha = \mathfrak{\tilde{c}}\mathfrak{c}\,e^{2\alpha\varphi}$ with $\alpha \in \mathbb{C}$. On the sphere for $h=0$ the conformal bootstrap has determined the three-point structure constants for the Liouville vertex operators given by the Dorn-Otto-Zamolodchikov-Zamolodchikov (DOZZ) coefficients $\mathcal{C}(\alpha_1,\alpha_2,\alpha_3)$ \cite{DO, ZZ}. Referring to these references for the expression for general $\alpha_i$ we only report the structure constant for the identity operator $\mathcal{V}_b=\mathfrak{\tilde{c}}\mathfrak{c}\,e^{2b\varphi}$
\begin{equation}\label{Cbbb_appendix}
\mathcal{C}(b,b,b) = -\Lambda^{\frac{1}{b^2}-2}(\pi\gamma(b^2))^{Q/b}\frac{(1-b^2)^2}{\pi^3b^5 \gamma(b^2)\gamma(b^{-2})}e^{-Q^2+ Q^2\log 4}~,
\end{equation}
where $\gamma(x)\equiv \Gamma(x)/\Gamma(1-x)$.

To avoid a conformal anomaly in (\ref{Zgrav2D_weyl}) we need the condition $c_{\text{m}}+ c_L-26=0$ where $c_{\text{m}}$ is the central charge of the matter theory and $c_{\text{gh}}=-26$ the ghost central charge arising when we fixed the Weyl gauge. This constraint leads to
\begin{equation}
Q= \sqrt{\frac{25-c_{\text{m}}}{6}}~,\quad b= \frac{1}{2\sqrt{6}}\left(\sqrt{25-c_{\text{m}}}- \sqrt{1-c_{\text{m}}}\right)~,
\end{equation}
from which we infer that the reality conditions of $Q$ and $b$ depend on the value of the matter central charge $c_{\text{m}}$.
$S_L$ is real valued only for $c_{\text{m}}\leq 1$. For $c_{\text{m}} \in (1,25)$, the action is intrinsically complex, and new methods are required to deal with it. Above $c_{\text{m}}=25$ both $Q$ and $b$ are imaginary. We can Wick rotate $\varphi \rightarrow \pm i \varphi$, $b\rightarrow \mp i\beta$, $Q\rightarrow \pm i q$, such that the newly obtained Liouville action is again real valued. It is known as timelike Liouville theory and has the action
\begin{equation}
S_{tL}[\varphi]= \frac{1}{4\pi} \int_{\Sigma_h} \dd x^2 \sqrt{\tilde{g}}\left(-\tilde{g}^{ij}\partial_i \varphi \partial_j\varphi -q \tilde{R}\varphi + 4\pi \Lambda e^{2\beta\varphi}\right)~.
\end{equation}
The wrong sign kinetic term in timelike Liouville theory is analogous to the  conformal mode problem of the  higher dimensional Euclidean gravitational path integral \cite{GHP}. Timelike Liouville is conjectured to be a two-dimensional conformal field theory with central charge $c_{tL}= 1-6q^2$ obeying the constraint $c_{tL}+ c_{\text{m}} -26=0$. For the identity operator the structure constant $\mathcal{C}_{tL}$ on the sphere is given by the analytic continuation of (\ref{Cbbb_appendix})
\begin{equation}
\mathcal{C}_{tL}(\beta,\beta,\beta) =\pm i\Lambda^{-\frac{1}{\beta^2}-2}\left(\pi \gamma(-\beta^2)\right)^{-\frac{1}{\beta^2}+1}\frac{(1+\beta^2)}{\pi^3 \beta^5 \gamma(-\beta^2)\gamma(-\beta^{-2})}\,e^{q^2 -q^2 \log 4}~,
\end{equation}
as revealed by the semiclassical treatment in \cite{timelike,Muhlmann:2022duj,Giribet:2022cvw}.

\section{Dirichlet problem on timelike boundary}\label{IBVP}

We are interested in the development of solutions to the Einstein equations
\begin{equation}
R_{\mu\nu} - \frac{1}{2} g_{\mu\nu} R + \Lambda g_{\mu\nu} = 0~,
\end{equation}
on a manifold of the form $\mathcal{M} = I \times S$, with $I$ being a finite interval and $S$ a spatial three-manifold with boundary $\Sigma$. The boundary of the manifold is a timelike hypersurface $\mathcal{C} = I \times \Sigma$.  For simplicity, we consider the problem in the absence of matter fields. We will consider the problem at the linearised level, as adapted from \cite{An:2021fcq,Witten:2018lgb}. The deviation of a classical solution $g_0$ is denoted by $h_{\mu\nu}$ {such that $g_{\mu\nu}= g_{0,\mu\nu}+ h_{\mu\nu}$}. 

In a local region around a point on $\mathcal{C}$, the metric $g_{0,\mu\nu}$ looks like a portion of Minkowski space whose coordinates we take to be
\begin{equation}
ds^2 = -\dd t^2 + \dd x^2 +\dd y^2 + \dd z^2~,
\end{equation}
with left boundary at $x=0$ (such that $x \ge 0$) and $t \in (0,L)$ for some length scale $L$ which is taken to be one. In this local region, the Dirichlet boundary conditions on $h_{\mu\nu}$ correspond to\footnote{We use the notation that the $|$ corresponds to the value of a field at $x=0$.} $h_{mn}| = 0$, where $m,n = \{t,y,z\}$. We must also impose conditions on the induced metric and extrinsic curvature along the spacelike surface at $t=0$. Finally, linearised diffeomorphisms must not disrupt the Dirichlet conditions. The linearised gauge parameter $\xi_\mu$ acting on $h_{\mu\nu}$ must satisfy $\xi_m = 0$ so as not to affect the condition $h_{mn}| = 0$, as well as $\xi_x = 0$ so as not to disrupt the location of the boundary itself. Consequently, $\xi_\mu | = 0$. 

As a gauge fixing condition, we pick the de Donder gauge 
\begin{equation}\label{dedonder}
\nabla^\nu h_{\mu\nu} - \frac{1}{2} \partial_\mu {h^\nu}_\nu = 0~.
\end{equation}
Implementing the de Donder gauge condition imposes further restrictions on $h_{x m}|$ and $h_{xx}|$. Near the Minkowski region we can expand the solutions in terms of superpositions of plane waves
\begin{equation}\label{planewave}
h^{(\bold{k})}_{\mu\nu}(x^\mu) = \gamma_{\mu\nu}(\bold{k}) \, e^{i k_\mu x^\mu}~, \quad\quad  k^\mu k_\mu \equiv -(k^0)^2 +\bold{k}^2= 0~,
\end{equation}
where $\gamma_{mn} = 0$ by the Dirichlet condition. 
The trace of $\gamma_{\mu\nu}$ is denoted by $\gamma \equiv {\gamma_{xx}}$. Imposing the condition (\ref{dedonder}) further restricts 
\begin{equation}
k_x \gamma_{mx} - \frac{1}{2}  k_m \gamma = 0~, \quad\quad  k^m \gamma_{mx} + \frac{1}{2}  k_x \gamma = 0~.
\end{equation}
The above solutions are both satisfied by
\begin{equation}
\gamma_{m x}(\bold{k}) = \frac{k_m}{2 k_x}\gamma (\bold{k})~.
\end{equation}
This $\gamma_{\mu\nu}$ is not fully fixed by the Dirichlet boundary condition. We must also impose boundary conditions on the initial spacelike surface $t=0$ and $x \ge 0$. As a particular example, we impose that $h_{\mu\nu} = \partial_t h_{\mu\nu} = 0$. The general solution is a superposition of modes 
\begin{equation}\label{fourier}
h_{\mu\nu}(x^\mu) = \int_{\mathbb{R}^3} \frac{\dd^3\bold{k}}{(2\pi)^3} \gamma_{\mu\nu}(\bold{k}) e^{-i |\bold{k}| t + i \bold{k} \cdot \bold{x}} + \text{h.c.}~,
\end{equation}
All that remains is to adjust the $\gamma_{\mu\nu}(\bold{k})$ to satisfy the conditions on $t=0$ and $x \ge 0$. As a simple example, we can take $k_y = k_z = 0$ and further restrict $k_x \ge 0$ such that 
\begin{equation}
\gamma_{m x}(\bold{k})  =  \frac{\delta_{m t}}{2}\gamma(\bold{k})~. 
\end{equation}
Such a theory has purely right-moving waves. We now need to understand whether there exist such waves satisfying $h_{\mu\nu} = \partial_t h_{\mu\nu} = 0$ at $t= 0$ and $x\ge 0$. The non-vanishing components of $h_{\mu\nu}$ follow from (\ref{fourier}) and are given by
\begin{equation}\label{superp}
h_{xx}(x^\mu) =2 h_{tx}(x^\mu) =  \int_{\mathbb{R}^+} \frac{\dd {k_x}}{2\pi} \left( \gamma(k_x) e^{- i k_x (t-x)} + \gamma^*(k_x) e^{i k_x (t-x)}   \right) \equiv f_\gamma(t-x)~.
\end{equation}
To satisfy the remaining boundary conditions $h_{\mu\nu} = \partial_t h_{\mu\nu} = 0$ at $t = 0$ and $x \ge 0$, we need to construct a function $f_\gamma(w)$ such that $f_\gamma(w) = f'_\gamma(w) = 0$ for $w \le 0$. For example, we can take $f_\gamma(w)$ to be the bump function 
\begin{equation}
f_\gamma(w) = \alpha_\gamma \, e^{-\left({1-(5-w)^2}\right)^{-1}}~, \quad\quad  w \in (4,6)~,
\end{equation}
while vanishing everywhere else. The parameter $\alpha_\gamma \in \mathbb{R}$ is a constant indicating the height of the bump function. The above function is infinitely differentiable and has support entirely on the negative real $w$-axis. The function $f_\gamma(w)$ is square integrable along the real axis and thus has a well defined Fourier transform, it can be built by a superposition (\ref{superp}) of plane waves. Moreover at $t=0$ the function $f_\gamma(t-x)$ as well as its time derivative vanish at positive values of $x$. Given that the configuration $h_{\mu\nu}=0$ is also a solution of the linearised Einstein equations that  satisfies all the boundary conditions, we see that the Dirichlet problem under consideration does not have a unique solution. Similar reasoning holds for the Neumann case. 

We note that the linearised solution (\ref{superp}) is locally pure gauge, it has vanishing Riemann tensor. Indeed, we can express it as $h_{\mu\nu} = \partial_\mu \xi_\nu + \partial_\nu \xi_\mu$~, with
\begin{equation}
\xi_x(x^\mu) = - \frac{1}{2} \int^{t-x}_{-\infty} d u  \, f_{\gamma}(u)~, \quad\quad \xi_m = 0~.
\end{equation}
Consequently, $\xi_\mu$ does not vanish along the timelike hypersurface $\mathcal{C}$ in contradiction to the boundary conditions imposed on the gauge parameter. Therefore, the solution (\ref{superp})  cannot be gauged away and is instead physical, it is reminiscent of the large gauge transformations considered in the context of asymptotically Minkowski space.

In \cite{An:2021fcq} it is also suggested that fixing the conformal class of the boundary metric and the trace of the exstrinsic curvature may lead to a well-posed problem. At the linearised level, and working about a flat background as before, the conformal class condition fixes instead that $\gamma_{mn} = \Gamma(\bold{k}) \, \eta_{mn} $ for some function $\Gamma(\bold{k})$.  Fixing the trace  $K = {K^\mu}_\mu$ of the extrinsic curvature $K_{\mu\nu} = \nabla_\mu n_\nu$, with $n^\mu$ a unit vector normal to the timelike hypersurface $\mathcal{C}$, leads to the following condition at the linear level 
\begin{equation}
 \frac{1}{2} \partial_x {h_m}^m - \partial^m h_{mx} = 0~.
\end{equation}
Once again, imposing the de Donder gauge (\ref{dedonder}) and taking the solutions to be superpositions of plane waves (\ref{planewave}) we find the conditions 
\begin{eqnarray}
k^m \gamma_{mx} -   \frac{3k_x}{2}  \Gamma &=& 0~, \\
k^m \gamma_{mx} + \frac{k_x}{2} \gamma_{xx} -  \frac{3}{2} k_x  \Gamma &=& 0~, \\
k_x \gamma_{mx} -\frac{k_m}{2} \left(\gamma_{xx} + \Gamma \right) &=& 0~.
\end{eqnarray}
Provided $k_x \neq 0$, the above equations are solved for $\Gamma = \gamma_{mx} = \gamma_{xx} = 0$. For $k_x = 0$, one has the more general solution
\begin{equation}
k^m \gamma_{mx} = 0~, \quad\quad \gamma_{xx} = -\Gamma~.
\end{equation}
We must recall, however, that we must impose a condition at $t=0$ and $x\ge 0$. Let us impose, as before, that $h_{\mu\nu} =  \partial_t h_{\mu\nu} = 0$ along $t=0$ and $x\ge 0$. For $h_{xx}$ the most general superposition with $k_x=0$ reads
\begin{equation}
h_{xx}(x^\mu) = \int_{ \mathbb{R}^2} \frac{\dd k_y \dd k_z}{(2\pi)^2}  \gamma_{xx}(k_y,k_z)  e^{-i \sqrt{k_y^2+k_z^2} \, t + i k_y y + i k_z z} + \text{h.c.}~.
\end{equation}
Imposing that $h_{x x} =  \partial_t h_{xx} = 0$ along $t=0$ and $x\ge 0$ immediately leads to $\gamma_{xx} = -\Gamma  = 0$. Similarly, we conclude that $\gamma_{m x} = 0$. 

Interestingly, the conformal boundary conditions of \cite{An:2021fcq} have appeared previously in the fluid gravity literature \cite{Bredberg:2011xw,Anninos:2011zn}. It was noted in these works that there is an obstruction in solving the Einstein equations for a timelike boundary endowed with Dirichlet boundary conditions parameterically near the event horizon of a black hole \cite{Bredberg:2011xw} or the de Sitter horizon \cite{Anninos:2011zn}. It is noted in the same work, that the obstruction is lifted when fixing instead the conformal class of the induced metric and trace of the extrinsic curvature of the timelike boundary, as in \cite{An:2021fcq}.

\section{Flow geometries}\label{app:flow}
In this appendix we provide some details for the flow geometry (\ref{flow_geometry}). 
Following \cite{Centaur3} we take a family of static, spherical symmetric spacetimes with metric
\begin{equation}
ds^2 = -f(\rho)\dd t^2 + \frac{\dd \rho^2}{f(\rho)}+ g(\rho)^2\dd \Omega_2^2~,
\end{equation}
where $f(\rho)$ and $g(\rho)$ are continuous functions satisfying $g(\rho_h)\neq 0$ and $f(\rho_h)=0$ at the horizon $\rho= \rho_h$ and $f(\rho) > 0$, $g(\rho) >0$ for $\rho>\rho_h$. The Einstein equation relate the functions $f(\rho)$ and $g(\rho)$ to the matter stress tensor $T_{\mu\nu}$, which we assume satisfies the null energy condition $T_{\mu\nu}v^\mu v^\nu\geq 0$ for future pointing null vectors $v^\mu$. This in turn imposes the constraints
\begin{equation}\label{eq:inequalities}
f(\rho) g(\rho) g''(\rho)\geq 0~,\quad 1-g'(\rho)^2 f(\rho)+ \frac{1}{2}g(\rho)^2 f''(\rho) - f(\rho) g(\rho) g''(\rho) \geq 0~.
\end{equation}
We are now deriving a set of functions $f(\rho)$ and $g(\rho)$ that produce a geometry interpolating from AdS$_2\times S^2$ to the static patch horizon of dS$_4$ while satisfy the null energy constraints (\ref{eq:inequalities}).
The metric of the static patch of dS$_4$ is given by
\begin{equation}
ds^2 = - \left(\rho(2\ell-\rho)\right) \frac{\dd t^2 }{\ell^2}+ \frac{\ell^2}{\rho(2\ell-\rho)} \dd \rho^2 + \ell^2\left(1-\frac{\rho}{\ell}\right)^2 \dd \Omega_2^2~,
\end{equation}
where the horizon is at $\rho_h =0$. 
\newline\newline
\textbf{Example I.}
One example with the required properties is given by
\begin{equation}
%
f(\rho) =  \tanh\left(\frac{2\rho \ell -\rho^2}{\ell^2}\right)+ \left(\frac{\rho}{\ell_{\scaleto{\text{AdS}}{4.5pt}}}\right)^{3-\frac{1}{a}}\tanh\left(\left(\frac{\ell_{\scaleto{\text{AdS}}{4.5pt}}}{\rho}\right)^{1-\frac{1}{a}}\right)~,\quad g(\rho) = \ell-\rho~, \quad a>1~.
\end{equation}
For $\ell_{\scaleto{\text{AdS}}{4.5pt}} \ll \ell$, we can view the geometry as an interpolation from an asymptotically  AdS$_2\times S^2$ geometry in the region a to the static patch of dS$_4$  in region b,
\begin{equation}
\text{a.}\quad \left(\frac{\ell_{\scaleto{\text{AdS}}{4.5pt}}}{\ell}\right)^{\frac{a-1}{2a-1}} \ll \frac{\rho}{\ell}\lesssim 0~,\qquad \text{b.}\quad \left(\frac{\ell_{\scaleto{\text{AdS}}{4.5pt}}}{\ell}\right)^{\frac{a-1}{2a-1}} \gg \frac{\rho}{\ell}>0~,\quad\quad a>1~.
\end{equation}
\textbf{Example II.}
We again fix $g(\rho) = \ell - \rho$, which trivially saturates the first inequality. 
The most general AdS$_2$ metric can be parameterised by two parameters, $\rho_*$ and $\alpha$, and is given by
\begin{equation}
f(\rho)_{\scaleto{\text{AdS}}{4.5pt}} = \left( \frac{\rho - \rho_*}{\la} \right)^2 + \alpha \,.
\end{equation}
We want to find a function $f(\rho)$ that satisfies the second inequality in (\ref{eq:inequalities}) and has the following two properties. First, for $\rho_h \leq \rho \lesssim \rho_c$, it behaves as $f(\rho) = f(\rho)_{\scaleto{\text{dS}}{4.5pt}} \equiv  (2\rho \ell - \rho^2)/\ell^2$, for some $\rho_h<\rho_c<\ell$. Second, for $\rho_c \lesssim \rho$, it behaves approximately as $f(\rho)_{\scaleto{\text{AdS}}{4.5pt}}$. For simplicity, we choose $\rho_* = \rho_c$ and $\alpha = 1$. A smooth function that does the job is given by
\begin{equation}
f(\rho) = \frac{2\rho \ell - \rho^2}{\ell^2} + \left( \left( \frac{\rho - \rho_c}{\la} \right)^2 + 1 \right) \left(  \frac{1 + \tanh \frac{\rho-\rho_c}{\varepsilon}}{2}\right) \,,
\end{equation}
provided that $\ell/\la \gg 1$. The parameter $\varepsilon$ controls the width of the interpolating region. When $ \varepsilon \sim \la$, this function satisfies the null energy condition up to some $\rho_b \lesssim \ell$. Note that $\rho_b/\la$ will be parametrically large, and thus near the boundary of AdS$_2$.

 \begingroup
 \addcontentsline{toc}{section}{References}

\end{document}